\documentstyle[aps,prb,amsfonts,amsmath,amssymb,epsfig,graphicx]{revtex}
\def\minifrac#1#2{{\textstyle\frac{#1}{#2}}}
  \def\nz{{\mathbb N}}
  
  \def\Tr{\text{Tr}}
  
  \def\Im{\text{Im}}
  \def\e{{\rm e}}
  \def\d{{\rm d}}
  
  \def\O{{\cal O}}
 \def\defeq{{\mathchoice{%
    \mathrel{\mskip\thickmuskip\raise.4pt\hbox{$\mathord{\displaystyle:}$}%
    \hbox{$\mathord{\displaystyle=}$}\mskip\thickmuskip}}{%
    \mathrel{\mskip\thickmuskip\raise.4pt\hbox{$\mathord{\displaystyle:}$}%
    \hbox{$\mathord{\displaystyle=}$}\mskip\thickmuskip}}{%
    \mathrel{\mskip.25\thinmuskip\raise.25pt\hbox{$\mathord{\scriptstyle:}$}%
    \hbox{$\mathord{\scriptstyle=}$}\mskip.25\thinmuskip}}{%
    \mathrel{\mskip.1\thinmuskip\raise.1pt%
             \hbox{$\mathord{\scriptscriptstyle:}$}%
    \hbox{$\mathord{\scriptscriptstyle=}$}\mskip.1\thinmuskip}%
  }}}
  \def\eqdef{{\mathchoice{%
    \mathrel{\mskip\thickmuskip\hbox{$\mathord{\displaystyle=}$}%
    \raise.4pt\hbox{$\mathord{\displaystyle:}$}\mskip\thickmuskip}}{%
    \mathrel{\mskip\thickmuskip\hbox{$\mathord{\displaystyle=}$}%
    \raise.4pt\hbox{$\mathord{\displaystyle:}$}\mskip\thickmuskip}}{%
    \mathrel{\mskip.25\thinmuskip\hbox{$\mathord{\scriptstyle=}$}%
    \raise.25pt\hbox{$\mathord{\scriptstyle:}$}\mskip.25\thinmuskip}}{%
    \mathrel{\mskip.1\thinmuskip\hbox{$\mathord{\scriptscriptstyle=}$}%
    \raise.1pt\hbox{$\mathord{\scriptscriptstyle:}$}\mskip.1\thinmuskip}%
  }}}
  \let\Ds=\displaystyle

\newcommand{\myscalebox}[1]{\scalebox{0.6}[0.6]{#1}}
\newcommand{\myscaleboxb}[1]{\scalebox{0.8}[0.8]{#1}}

\begin{document}
\def\e#1{\exp\left\{#1\right\}}
\def\ehoch#1{{\rm e}^{#1}}
\def\Tr{\mathop{\rm Tr}\nolimits}
\def\average#1{\left\langle#1\right\rangle}
\def\drm{{\rm d}}
\newif\ifboo \boofalse

\def\And{{\rm and\ }}
\def\Review#1{\boofalse{\it #1},}
\def\Name#1{{\sc #1},}
\def\Vol#1{\ifboo Vol. {\bf #1}\else{\bf #1}\fi}
\def\Year#1{\ifboo #1\else(#1)\fi}
\def\Book#1{\bootrue{\it #1},}
\def\Page#1{\ifboo {\rm p. #1}\else{\rm #1}\fi}
\title{Stress Relaxation of Near-Critical Gels} 

\author{Kurt Broderix\footnote{deceased (2 May 2000)}, Timo
  Aspelmeier,  Alexander K. Hartmann and
Annette Zippelius}
\address{Institute for Theoretical Physics,
      University of G\"ottingen, Bunsenstr. 9, 37073 G\"ottingen, Germany}
\date{\today}
\maketitle

\begin{abstract}
The time-dependent stress relaxation for a Rouse model of a
crosslinked polymer melt is completely determined by the spectrum of
eigenvalues of the connectivity matrix. The latter has been computed
analytically for a mean-field distribution of crosslinks.  It shows a
Lifshitz tail for small eigenvalues and all concentrations below the
percolation threshold, giving rise to a stretched exponential decay of
the stress relaxation function in the sol phase. At the critical point
the density of states is finite for small eigenvalues,
 resulting in a logarithmic divergence
of the viscosity and an algebraic decay of the stress relaxation
function. Numerical diagonalization of the connectivity matrix supports
the analytical findings and has furthermore been applied to cluster
statistics corresponding to random bond percolation in two and three
dimensions. 
\end{abstract}

\section{Introduction}

The most striking observation in near critical gels is the anomalous stress
relaxation \cite{Martin91}, which precedes the transformation of the viscous
fluid into an elastic amorphous solid, i.e. the gelation transition.  Here,
polymer systems are considered, where the viscoelastic behavior is controlled
by the concentration $c$ of crosslinks connecting monomers of different
molecules. At a critical concentration $c_{\text{crit}}$ the gelation
transition occurs.  Viscoelastic studies by several groups have revealed the
following characteristic features of stress relaxation:
\begin{itemize}
\item In the sol phase, well below the gelation transition, one
  observes a stretched exponential decay of the stress relaxation
  function $\chi(t) \sim \exp{-(t/\tau^*)^{\beta}}$.
\item The time scale $\tau^* \sim \epsilon^{-z}$ diverges as the critical
  point is approached. Here $\epsilon=(c_{\text{crit}}-c)/c_{\text{crit}}$ denotes the distance from the
  critical point.
\item The viscosity $\eta$, which is given as the integral over the
  stress relaxation function, diverges $\eta \sim \epsilon ^{-k}$ as
  the critical point is approached.
\item At the critical point, stress relaxation is algebraic in time: $\chi(t)
  \sim t^{-\Delta}$.
\end{itemize}

Whereas the stretched exponential decay is characteristic of the sol phase and
holds for all crosslink concentrations $c<c_{\text{crit}}$, the last three
observations refer to critical behavior, as the gel point is approached. If
dynamic scaling holds, these findings can be cast in a scaling ansatz for the
stress relaxation function $\chi(\epsilon,t)$, which depends on time and
distance from the critical point $\epsilon$:
\begin{equation}
\label{Eq1}
\chi(c,t)=\epsilon^{z-k}g(t/\tau^*(\epsilon))
\end{equation}
with $\tau^* \sim \epsilon^{-z}$.  Given a certain distance $\epsilon$ from
the gel point, one expects to see a crossover from an algebraic decay at
intermediate times to a stretched exponential decay at asymptotically large
times.  The scaling ansatz implies $\Delta=(z-k)/z$.  Dynamic scaling as
implied by Eq.\ (\ref{Eq1}) is well confirmed experimentally \cite{Adolf90}
for the intermediate time regime where $\chi(t)$ decays like a power
law. However, the values for the exponents scatter considerably. Martin et
al.\ \cite{Martin88} and Adolf et al.\ \cite{Adolf90} find $\Delta =0.7\pm
0.05$ in agreement with the value $\Delta =0.7\pm 0.02$ of Durand et al.\
\cite{Durand87}, whereas Winter and coworkers \cite{Winter87} observe a wide
range of exponent values $0.2 \leq \Delta \leq 0.9$, depending on molecular
weight and stoichiometry. The experimental support for a universal stretched
exponential law is weak. Whereas Martin et al. confirm the stretched
exponential decay and quote $\beta\sim 0.4$ \cite{Martin88}, other studies
reveal non-universal exponents $\beta$.  The divergence of the time scale
$\tau^*\sim \epsilon ^{-z}$ in the scaling function was determined in
viscoelastic measurements as $z=3.9\pm0.2$ \cite{Martin88,Adolf90} and deduced
from static measurements of the shear modulus as $z=4.0\pm0.6$
\cite{Adam85}. The experimental values for k, the critical exponent of the
viscosity, vary in the range $0.7 \leq k \leq 1.4$. The origin of the scatter
in the experimental data is not clear.  One possible explanation is the size
of the critical region, which is known to depend on the degree of
polymerization.  Hence experiments with different samples have to cope with
different sizes of the critical region and possibly strong crossover effects.

In this paper we study the simplest dynamic model -- Rouse dynamics -- in the
presence of a time dependent shear flow by means of analytical calculations
and numerical simulations. Within this model, the frequency dependent stress
relaxation is completely determined by the spectral properties of the
connectivity matrix $\Gamma$, which specifies which monomers are
crosslinked. As a function of the total concentration of crosslinks $c$ one
observes in general a percolation transition at a critical concentration
$c_{\text{crit}}$, such that for $c<c_{\text{crit}}$ no macroscopic cluster of
connected particles exist, whereas for $c>c_{\text{crit}}$ the system
percolates. In the context of gelation the fraction of sites in the
macroscopic cluster has been identified with the gel fraction and the
percolation transition has been shown to mark the onset of solidification
\cite{Goldbart96}.

The connectivity matrix $\Gamma$ is a positive semidefinite, random matrix,
which has been studied in various contexts, e.g. diluted ferromagnets,
diffusion in sparsely connected spaces \cite{Bray88}, anomalous relaxation in
glassy systems and localization \cite{Biroli99}. In all cases the system
consists of $N$ vertices (monomers in the context of gelation) which are
connected by $cN$ edges (crosslinks). In the simplest case (mean field) one
chooses the edges independently out of all possible $N(N-1)/2$ edges. The
density of eigenvalues can be computed analytically for the above simple
distribution and has been discussed in Refs.\ \onlinecite{Bray88,Biroli99} in
the percolating regime, i.e.\ $c\geq c_{\text{crit}}$.  In this paper we focus
on the range $c \leq c_{\text{crit}}$, which corresponds to the sol phase and
the critical point.  For crosslinks of unit strength the spectrum of $\Gamma$
is shown to consist of $\delta$-functions only.  For fluctuating crosslink
strength, the spectrum is smooth and shows a Lifshitz tail for small
eigenvalues. The spectrum determines the time dependent stress relaxation
function of Eq.\ (\ref{Eq1}). In mean field theory the exponents are found to
be $\beta=1/3, \Delta=1$ and $z=3$. These results have been confirmed by
numerical diagonalisation of the connectivity matrix $\Gamma$.

The latter approach can be extended to finite dimensional connectivities,
corresponding to two and three dimensional percolation. Qualitatively the
results are similar to mean-field theory with, however different exponent
values. In $d=2$ we find $\Delta \approx 0.74$ and $k\approx 1.19$. The latter
value is in good agreement with a scaling relation, which was derived
previously \cite{Broderix99} and yields $k\sim 1.17$ using high precision data
for the conductivity exponent \cite{Grassberger1999}. Our numerical analyis in
$d=3$ is restricted to rather small system sizes ($\leq 20^3$) as compared to
high precision data of Gingold et al.\ \cite{Gingold1990}, because these
authors restrict their investigations to the conductance, whereas we aim at
the density of states, which requires a much larger numerical
effort. Consequently our exponent values are of relatively low precision as
compared to Ref.\ \onlinecite{Gingold1990}. Specifically we find
$\Delta\approx 0.83$ and $k\approx 0.75$, whereas scaling together with high
precision data yield for the latter $k\sim 0.71$.

The paper is organised as follows: In the following section (Sec.\
\ref{sec:two}) we introduce the dynamic model and the observables which we
want to discuss and which can be related to the spectrum of eigenvalues of the
connectivity matrix. In Sec.\ \ref{sec:three} we present the analytical
calculations for the mean-field distribution of crosslinks. We briefly review
the derivation of a selfconsistent equation for the spectrum, which was
previously given by Bray and Rodgers \cite{Bray88}. We then go on to discuss
the appearance of Lifshitz tails for small eigenvalues. For crosslinks of unit
strength the spectrum is shown to consist of a countable set of
$\delta$-peaks. We present an analytical scheme to systematically compute the
spectrum by iteration.  We also consider crosslinks of fluctuating stength,
for which the spectrum is continuous and can be obtained by standard numerical
means from the selfconsistent integral equation. In Sec.\ \ref{sec:four} we
present results from a numerical diagonalisation of random connectivity
matrices. We first compute the spectrum for a mean-field distribution of
crosslinks and compare it to the analytical results. Next, cluster
distributions of random bond percolation in 2 and 3 dimensions are
considered. Data for the stress relaxation function is presented as well as
finite size scaling plots for the static shear viscosity. We summarize our
results in Sec.\ \ref{sec:five}. Some detailed calculations have been deferred
to appendices.

Our paper is an extension of previous work, in which we discussed the static
shear viscosity \cite{Broderix99,Broderix2000} and self diffusion
\cite{BrGoZi97} in the sol phase as well as at the gelation transition. There
it was shown that the long time limit of the incoherent scattering function is
determined by the zero eigenvalues of the connectivity matrix, and the static
shear viscosity is determined by the trace of the Moore-Penrose inverse of the
connectivity matrix. Here we focus on the {\it full} spectrum of eigenvalues,
which also determines the decay of the stress relaxation at {\it finite}
times.
 
\section{Model and Observables}
\label{sec:two}

We consider a system of $N$ identical Brownian particles, each characterized
by its time-dependent position vector ${\bf R}_i(t)$ ($i=1,\ldots,N$) in
$d$-dimensional space of volume $V$, i.e.\ with density $\rho=N/V$. $M$
permanent crosslinks are introduced between randomly chosen pairs of particles
$(i_e,i'_e)$, resulting in a crosslink concentration $c=M/N$. These crosslinks
are modelled by a harmonic potential
\begin{equation} \label{Eq4}
  U := \frac{d}{2a^2}\:\sum_{e=1}^M \lambda_e
  \bigl( {\bf R}_{i_e}-{\bf R}_{i'_e} \bigr)^2,
\end{equation}
whose overall strength is controlled by the parameter $a>0$.  We use units of
energy such that $k_{\text{B}}T=1$ and allow for fluctuations in the strength
of crosslinks by introducing the parameter $\lambda_e$. Crosslinks of uniform
strength correspond to all $\lambda_e=1$. In general each crosslink $e$ is
assigned independently a random strength $\lambda_e$ according to the
distribution $p(\lambda)$.  The connectivity of the particles is specified by
the connectivity matrix
\begin{equation}
   \label{Eq4_1}
      \Gamma_{ii'}     = \sum_{e=1}^{M} \lambda_e
    (\delta_{ii_{e}} - \delta_{ii_{e}'})
    (\delta_{i'i_{e}}- \delta_{i'i_{e}'})\, , 
\end{equation}  
in terms of which the potential reads $U= (d/2a^{2})\sum_{i,i'=1}^{N}
\Gamma_{ii'}\,
 {\bf R}_i\cdot {\bf R}_{i'}$.  As usual
$\delta_{ij}$ denotes the Kronecker symbol, i.e.
$\delta_{ij}=1$ for $i=j$ and zero otherwise.  

We consider purely relaxational dynamics in an externally applied
space- and time-dependent velocity field $v^{\alpha}_{{\rm
    ext}}({\bf r},t)$:
\begin{equation} \label{Eq2}
  \partial_t R^{\alpha}_i(t) = 
  - \frac{1}{\zeta} \: \frac{\partial U}{\partial R^{\alpha}_i}(t) 
  + v^{\alpha}_{{\rm ext}}\bigl({\bf R}_i(t),t\bigr) + \xi^{\alpha}_i(t).
\end{equation}
 Here, Greek indices indicate Cartesian
co-ordinates $\alpha=x,y,z,\dots$, and we will always consider a flow
field in the $x$-direction, increasing linearly with $y$, {\it i.e.}
\begin{equation}
  \label{Eq1_1}
  v^{\alpha}_{{\rm ext}}({\bf r},t) =
  \delta_{\alpha,x}\kappa(t)r_y,
\end{equation}
with a time-dependent shear rate $\kappa(t)$.  The noise
${\boldsymbol \xi}$ has zero mean and covariance
$\langle\xi^{\alpha}_i(t)\,\xi^{\beta}_{i'}(t')\rangle = {2}{\zeta}^{-1}\,
\delta_{\alpha,\beta}\, \delta_{i,i'}\, \delta(t-t')$, where $\delta(t)$ is
the Dirac $\delta$-function. Here, the bracket $\langle\cdots\rangle$
indicates the average over the realizations of the Gaussian noise
$\boldsymbol\xi$.  The relaxation constant is denoted by $\zeta$.  The
probability distribution of crosslink configurations ${\cal
  G}=\{i_e,i'_e\}_{e=1}^M$ as well as the probability distribution of
crosslink strengths will be specified later.

In Ref.\ \onlinecite{Broderix99} we have 
computed the shear viscosity in the sol phase for (macro-) molecular units of
arbitrary internal connectivity. It was shown that the dependence on the
crosslink concentration and in particular the critical behavior near the
gelation threshold is the same for all (macro-) molecular units, as long as we
consider identical units with a finite degree of polymerization. We expect the
same universal behavior for stress relaxation on long time scales, which are
much larger than the longest internal time scale of a single (macro-)
molecule.  Hence we specialize to the simplest units, namely Brownian
particles.

\subsection*{Relaxation of shear stress}

We aim at the computation of the intrinsic shear stress
$\sigma_{\alpha\beta}(t)$ as a function of the shear rate $\kappa(t)$.
For the simple shear flow (\ref{Eq1_1}), a linear response relation
\begin{equation} \label{Eq6_1}
  \sigma_{xy}(t)=
  \int_{-\infty}^t\drm\tau\:
  \chi(t-\tau)\kappa(\tau)
\end{equation}
is valid for arbitrary strengths of the shear rate $\kappa(t)$.  The linear
response or shear relaxation function $\chi(t)$ is given in terms of the
connectivity matrix as explained in detail in Ref.\ \onlinecite{Broderix2000}
\begin{equation}
\label{eq:stressrelax}
\chi(t)=\frac{\rho}{N}\sum_{i=1}^N \left((1-E_0({\cal G})) 
\e{-\frac{2dt}{\zeta a^2}\Gamma({\cal G})}\right)_{ii} =:
\frac{\rho}{N}\Tr \left((1-E_0({\cal G})) 
\e{-\frac{2dt}{\zeta a^2}\Gamma({\cal G})}\right).
\end{equation}
The matrix $E_0$ denotes the projector onto the subspace of zero
eigenvalues of $\Gamma$ \cite{Broderix99}.  For a time independent
shear rate, $\kappa(t)=\kappa$, the stress tensor is time independent
and related to the shear rate $\sigma=\rho \eta \kappa$ via the static
shear viscosity, given by \cite{Broderix99}
\begin{equation}
\eta = \frac{\zeta a^2}{2dN}\Tr\frac{1-E_0({\cal G})}{\Gamma({\cal G})}.
\end{equation}

\subsection*{Selfdiffusion}
To discuss selfdiffusion we set the externally applied
velocity field to zero and focus on the incoherent scattering function
\begin{equation}
  S({\bf q},t):=\lim_{T\to\infty}
  \left\langle
    \frac{1}{N}
    \sum_{i=1}^N 
\e{{\rm i}\,{\bf q}\,\bigl({\bf R}_i(t+T)-{\bf R}_i(T)\bigl)}\right\rangle
\end{equation}
and the squared time delayed displacement
\begin{equation}
C(t) := \lim_{T\to\infty} \left\langle  \frac{1}{N}
    \sum_{i=1}^N ({\bf R}_i(t+T)-{\bf R}_i(T))^2\right\rangle.
\end{equation}

We note that ${\bf R}_i(t+T)-{\bf R}_i(t)$ is a Gaussian Markov process whose distribution is
in the limit $T\to\infty$ characterized by a vanishing mean and the
covariance function
\begin{equation}\label{Eq20_03}
  \begin{split}
    G_{ii'}(t) &:=
    \lim_{T\to\infty}
    \left\langle
      \left[{\bf R}_i(t+T)-{\bf R}_i(T)\right]^{\phantom{1}}\!\!\!
      \left[{\bf R}_{i'}(t+T)-{\bf R}_{i'}(T)\right]\right\rangle\\
    &=\frac{1}{\zeta}
    \int_0^{t}\drm\tau\:
    \left(\e{-\frac{2d\tau}{\zeta a^2}\Gamma}\right)_{ii'}. 
  \end{split}
\end{equation}
Performing the integral in (\ref{Eq20_03}) leads to
\begin{equation}\label{Eq20_05}
  G_{ii'}(t)  =\frac{2}{\zeta}\left(\frac{\zeta a^2}{2d}\frac{1-E_0}{\Gamma}\:
 \left( 1-\e{-\frac{2dt}{\zeta a^2}\Gamma}\right)
    +t\:E_0\right)_{ii'}.
\end{equation}
The matrix $\Gamma$ is non-negative by inspection 
(see Eq.\ (\ref{Eq4})), 
as it should be to ensure relaxation to equilibrium.
The scattering function  as well as the time delayed displacement can
be expressed in terms of $G_{ii'}(t)$ via
\begin{equation}\label{Eq20_04}
  S({\bf q},t)= 
  \frac{1}{N}
  \sum_{i=1}^N
    \e{-q^2\:G_{ii}(t)}
\end{equation}
and
\begin{equation}
C(t)= \frac{1}{N}  \sum_{i=1}^N  G_{ii}(t).
\end{equation}
 
\subsection*{Density of eigenvalues}

All dynamic quantities of interest have been expressed in terms of
$\Gamma$. Accordingly, once we know the eigenvalues
$\{\gamma_i\}_{i=1}^N$ and eigenvectors of
this matrix, we can compute dynamic observables for arbitrary
times. In the following, we shall discuss the density of eigenvalues
\begin{equation}
\label{eq:densitydef}
D_{\text{tot}}(\gamma)=
  \lim_{N\to\infty}\overline{\frac{1}{N}\sum_{i=1}^{N}\delta(\gamma-\gamma_i)}
 = \lim_{N\to\infty}\overline{\frac{1}{N}\Tr\delta(\gamma-\Gamma)}
\end{equation}
for several crosslink distributions. Here $\overline{\,\cdot\,}$ denotes the
average over crosslink realizations. 
If one splits off the zero eigenvalues,
$D_{\text{tot}}(\gamma)$ can be written as
\begin{equation}
D_{\text{tot}}(\gamma)=T_0(c)\delta(\gamma) + (1-T_0(c))D(\gamma). 
\end{equation}
where $D(\gamma)$ is normalized to 1 and contains only the non-zero
eigenvalues.  If we group the particles into clusters, the eigenspace
of modes with zero eigenvalues corresponds to vectors that are
constant within one cluster \cite{BrGoZi97}. In other words there is
one zero eigenvalue for each cluster and the dimension of the null
space is just the number of clusters $N_{\text{cl}}$. The weight of
zero eigenvalues is simply given by the density of clusters, i.e.\ 
$T_0(c)=N_{\text{cl}}/N$.

We restrict ourselves to the density of eigenvalues and do not attempt
to compute eigenvectors, which is in general more difficult. Hence we
can only compute observables which can be written as 
$\frac{1}{N}\sum_{i=1}^{N}(f(\Gamma))_{ii}$, where $f$ is an
arbitrary function of $\Gamma$. The
incoherent scattering function is not of this form
(Eq.\ \ref{Eq20_04}), whereas the stress relaxation function is
\begin{equation}
\label{eq:stressrelax2}
\overline{\chi(t)}=(1-T_0(c))\rho\int_{0}^{\infty}\drm\gamma D(\gamma)
\e{-\frac{2dt}{\zeta a^2}\gamma}.
\end{equation}
The zero eigenvalues are not to be included in the integration, due to
the term $1-E_0$ in Eq.\ \eqref{eq:stressrelax}.  
Analogously, the averaged viscosity is given by
\begin{equation}
\label{eq:viscosity2}
\overline{\eta}=(1-T_0(c))\frac{\zeta a^2}{2d}\int_0^\infty\drm\gamma
\frac{D(\gamma)}{\gamma}.
\end{equation}
In the same way, the disorder averaged, time delayed displacement is
determined by
\begin{equation}
\overline{C(t)}=(1-T_0(c))\frac{a^2}{d}\int_{0}^{\infty}\drm\gamma
\frac{D(\gamma)}{\gamma}
\left(1-\e{-\frac{2dt}{\zeta a^2}\gamma}\right) + T_0(c) \frac{2t}{\zeta}.
\end{equation}
It can also be expressed as an integral over the time dependent response
function
\begin{equation}
\overline{C(t)}=\frac{2}{\zeta\rho}\int_0^{t}\drm\tau \overline{\chi(\tau)}+
T_0(c) \frac{2t}{\zeta}.
\end{equation}

\section{Mean field theory}
\label{sec:three}

We consider first the simplest distribution of crosslinks which ignores all
correlations between crosslinks, i.e.\ the crosslinks are chosen independently
of each other and each pair $(i_e,i'_e)$ of particle indices is realized with
equal probability. As shown in Ref.\ \onlinecite{ErRe60} the particle clusters
exhibit the analog of a percolation transition at a critical crosslink
concentration $c_{\text{crit}}=1/2$. Below this concentration there is no
macroscopic cluster and almost all finite clusters are trees. The average
number of tree clusters $T_n$ with $n$ particles is given in the macroscopic
limit by
\begin{equation}\label{Eq10_3}
\lim_{N\to\infty} \frac{T_n}{N}=  \tau_{n}=
\frac{n^{n-2}(2c\,{\rm e}^{-2c})^n}{2c\,n!}. 
\end{equation}
In particular the total number of clusters per particle is
\begin{equation}
T_0(c)=1-c.
\end{equation}
These results are independent of the distribution of crosslink
strengths, $p(\lambda)$. 

To compute the density of eigenvalues we introduce the resolvent
\begin{equation}
G(\Omega)=\lim_{N\to\infty}\frac{1}{N}\overline{\Tr \frac{1}{\Gamma-\Omega}}
\end{equation}
for complex argument $\Omega=\gamma + i\epsilon, \epsilon>0$. 
In the limit $\epsilon\to 0$, we recover the spectrum from the
imaginary part of the resolvent according to
\begin{equation}
D_{\text{tot}}(\gamma) = 
  \frac{1}{\pi}\lim_{\epsilon\downarrow 0} \Im G(\gamma+i\epsilon).
\end{equation}
It can be inferred from Eq.\ \eqref{eq:densitydef} that conversely,
$D_{\text{tot}}(\gamma)$ determines $G(\Omega)$ via
\begin{equation}
\label{eq:GfromD}
G(\Omega) = \int_{-\infty}^\infty\drm\gamma
  \frac{D_{\text{tot}}(\gamma)}{\gamma-\Omega}.
\end{equation}

\subsection{Disorder average by replicas}
Bray and Rodgers \cite{Bray88} have shown how to reduce the computation of
$D_{\text{tot}}(\gamma)$ for crosslinks of unit strength (i.e.\ all
$\lambda_e=1$) to the solution of a nonlinear integral equation.  Their
derivation is easily generalized to crosslinks of strength $\lambda$ which
fluctuates according to a given distribution $p(\lambda)$. We restrict
ourselves to distributions $p(\lambda)$ such that
\begin{equation} \label{Eq33}
  \int_0^\infty\frac{\d\lambda}\lambda\,p(\lambda) < \infty
\end{equation}
holds. It will be shown below (see Eq.\ (\ref{11})) that this condition is
necessary to ensure a finite viscosity in the sol phase. 
Following Bray and Rodgers we introduce a generating function 
\begin{equation}
Z(\Omega)=\int_{\mathbb{R}^N} \left(\prod_{i=1}^N
\frac{d\phi_i}{\sqrt{2\pi}}\right)
 \exp{\left(\frac{i}{2}\sum_{i,j} \phi_i\phi_j
    (\Omega \delta_{ij}-\Gamma_{ij})\right)},
\end{equation}
which determines the resolvent, according to
\begin{equation}
G(\Omega)=\lim_{N\to\infty}\frac{2}{N}\overline{\frac{\partial \log
    Z}{\partial \Omega}}.
\end{equation}
The average over the disorder is performed with the replica trick,
resulting in
\begin{equation}
\overline{Z^n}=\int_{\mathbb{R}^N}\left(\prod_{i=1}^N\prod_{\alpha=1}^n
\frac{d\phi_i^{\alpha}}{\sqrt{2\pi}}\right)
\exp{\left(\frac{i}{2}\Omega \sum_{i=1}^N\hat{\phi}_i\hat{\phi}_i 
+\frac{c}{N}\int_0^{\infty} d\lambda p(\lambda)\sum_{i,j=1}^N e^{-i\lambda 
(\hat{\phi}_i-\hat{\phi}_j)^2/2}
-cN\right)}.
\end{equation}
We assume that the connectivity is intensive, $\lim _{N\to\infty}(c/N)=0$ and
have introduced the notation 
$\hat{\phi}_i=(\phi_i^1,\phi_i^2,\hdots,\phi_i^n)$ for
n-times replicated variables. In the next step one decouples different sites
as shown in Ref.\ \onlinecite{Bray88} and performs a saddle point approximation for
large $N$. This gives rise to a self-consistent equation for a function
$g^{\Omega}(\hat{x})$
\begin{equation}
\label{Eq25}
g^{\Omega}(\hat{x})=2c\int_0^{\infty} d\lambda p(\lambda)\, 
\frac{\displaystyle\int d\hat{y} 
e^{(i\Omega \hat{y}^2+2g^{\Omega}(\hat{y})-i\lambda (\hat{y}-\hat{x})^2)/2}}
{\displaystyle\int d\hat{y} 
e^{(i\Omega \hat{y}^2+2g^{\Omega}(\hat{y}))/2}}
\end{equation}
which in turn determines the resolvent according to
\begin{equation}
G(\Omega)=\lim_{n\to 0}\frac{i}{n}\,\frac{\displaystyle\int d\hat{x}
\hat{x}^2 e^{(i\Omega \hat{x}^2/2+g^{\Omega}(\hat{x}))}}
{\displaystyle\int d\hat{x} e^{(i\Omega \hat{x}^2/2+g^{\Omega}(\hat{x}))}}.
\end{equation}
In the last step of the calculation one assumes a replica symmetric
solution for the saddle point equation: 
\begin{equation}
g^{\Omega}(\hat{x})=g^{\Omega}(\rho) \text{~with~} 
\rho=\sqrt{\sum_{\alpha}x^2_{\alpha}}.
\end{equation}
The limit $n\to 0$ can then be performed resulting in the following
nonlinear integral equation for $g^{\Omega}(\rho)$ (cf.\ Eqs.\ (16,17)
in Ref.\ \onlinecite{Bray88})
\begin{equation} \label{Eq35}
\begin{split}\Ds 
  g^{\Omega}(\rho)  & = \hbox{} \Ds
  2c\int_0^\infty\!\d\lambda\,p(\lambda)\,
  \exp\!\left\{-\frac{i\lambda}2\,\rho^2\right\} 
\\[3.5pt] \Ds & \Ds \hbox{} +
  2ic\, e^{-2c} \int_0^\infty\!\d\lambda\,p(\lambda)\int_0^\infty\!\d x\,
  \lambda\rho \,\text{I}_1(i\lambda\rho x)\,
  \exp\!\left\{
    -\frac{i\lambda}2\,(\rho^2+x^2)
    +\frac{i \Omega}2\,x^2 +g^{\Omega}(x) 
  \right\} 
\end{split}
\end{equation}
with $g^{\Omega}(0)=2c$.  Here $\text{I}_\nu(z)$ are the modified
Bessel functions of the first kind.  The solution of Eq.\ (\ref{Eq35})
yields the resolvent
\begin{equation} \label{Eq32}
  G(\Omega)  =- \int_0^\infty\frac{\d\lambda}\lambda\,p(\lambda) +
  \frac{i}{2c}\,\int_0^\infty\!\d\rho\,\rho\,g^{\Omega}(\rho) 
\end{equation}
and the density of eigenvalues
\begin{equation}
\label{eq:spectrum}
D_{\text{tot}}(\gamma)=\frac{1}{2c\pi}\lim_{\epsilon\to 0} 
\Im \left\{ i\int_0^{\infty} d\rho \rho g^{\gamma+i\epsilon}(\rho)\right\}.
\end{equation}

\subsection{Moments and Lifshitz tails}
If all inverse moments $M_n$ of the density of non-zero eigenvalues $ M_n
\defeq \int_{0}^\infty \d\gamma \gamma^{-n} D(\gamma)$, $n\in\nz $, exist,
one can derive the following asymptotic expansion of the resolvent
\begin{equation} \label{13}
  G(\Omega) = \frac{c-1}\Omega + \frac{2d\eta}{\zeta a^2} 
  + c\sum_{n=1}^\infty\Omega^n\,M_{n+1}
\end{equation}
by expanding the denominator in Eq.\ \eqref{eq:GfromD} in a geometric series.
As we show in Appendix A, the lowest moments are given explicitly by
\begin{equation} \label{11}
  M_1 = \frac 1{4c}  \left[\ln\left(\frac 1{1-2c}\right)-2c\right]
  \int_0^\infty\frac{\d\lambda}\lambda\,p(\lambda) = 
  \frac{2d\eta}{\zeta a^2}
\end{equation}
and
\begin{equation} \label{14}
  M_2=
  -\frac{\left(5\,P_2-4\,P_1^{2}\right)}{240c^2}\,
   \ln\left(\frac 1{1-2\,c}\right)
  -\frac{8\,{c}^{3}-6\,{c}^{2}-5\,c+1}{30c\left (1-2\,c \right )^{3}}
   \,P_1^2
  -\frac{4\,{c}^{2}-3\,c-1}{24c\left (1-2\,c\right )^{2}}
   \,P_2
\end{equation}
with $ P_n\defeq\int_0^\infty \d\lambda \lambda^{-n}\,p(\lambda)$.
Here and in the following we assume that very weak crosslinks are unlikely to
occur, because we are interested in the small eigenvalues which are due to the
geometry of the clusters and not due to the appearance of weak links. More
precisely we require
\begin{equation} \label{Eq24}
  \lim_{\lambda\downarrow 0} 
  \frac{\ln \left| \ln\, p(\lambda) \right|}
       {\left|\ln\lambda\right|} > \frac 12.
\end{equation}

The divergence of the moments, $M_1$ and $M_2$, suggests a Lifshitz tail of
the density of states of the following form
\begin{equation}\label{lifshitz}
D(\gamma) \propto \e{-\left(\frac{\gamma_0 (1-2c)^3}{\gamma}\right)^{\kappa}},
\quad \gamma\,\downarrow\,0,
  \quad c<\minifrac 12 ,
\end{equation}
since for positive $\kappa$, this ansatz implies for the inverse moments
\begin{equation}
M_n \propto (1-2c)^{-3(n-1)},\quad c\uparrow \minifrac 12 .
\end{equation}
Bray and Rodgers have given a heuristic argument in favor of the
ansatz (\ref{lifshitz})
with $\kappa=1/2$. They argue that out of all clusters for given $n$ the
linear one has the smallest eigenvalue, namely $\gamma_{\text{min}}=\gamma_0
n^{-2}$. There is just one linear cluster for given $n$, so that its
contribution to the spectrum is
\begin{equation}
D^{\text{lin}}=\frac{1}{2c}\sum_n \left(2ce^{(-2c)}\right)^n
\delta\left(\gamma-\frac{\gamma_0}{n^2}\right)
\sim e^{-\sqrt{\gamma_0/\gamma}}.
\end{equation}
Arbitrary finite clusters may be attached to the chain without altering
the dependence of the smallest eigenvalue on the length of the chain.
  If a finite cluster
of mass $m_i$ is attached to site i of the linear chain, the smallest
eigenvalue is $\gamma_{min}=\frac{\gamma_0}{m_i n^2}$. Replacing
\begin{equation}
m_i\sim \overline{m}=\frac{\sum_n n\tau_n}{\sum_n
  \tau_n}=\frac{1}{1-2c}
\end{equation} 
leads to $\gamma_{min}=\frac{\gamma_0 (1-2c)}{n^2}$. The number of
clusters contributing to $D(\gamma)$ for small $\gamma$ is much
larger, if attachments are taken into account: The probability to find
a chain of length $n$, regardless of attachments, is given by
$(2c)^n$. Hence the density of eigenvalues is estimated as
\begin{equation}
D^{\text{lin}}=\frac{1}{2c}\sum_n (2c)^n
\delta\left(\gamma-\frac{\gamma_0 (1-2c)}{n^2}\right)
\sim \exp\left\{-\left(\frac{\gamma_0 (1-2c)^3}{\gamma}\right)^{1/2}\right\}.
\end{equation}
Here, we have expanded $\ln(2c)\sim 2c-1$ for $c$ sufficiently close
to its critical value $c_{crit}=1/2$ to obtain the Lifshitz tail near
criticality. In Appendix A.3 we derive rigorous upper and lower
bounds for $D(\gamma)$, which prove that $D(\gamma)$ has indeed a
Lifshitz tail of the form $D(\gamma) \sim
\exp{(-\sqrt{h(c)/\gamma})}$. We are unable to derive the dependence
of $h(c)$ on crosslink concentration $c$, which is however suggested
by the lowest order moments (\ref{14}),(\ref{29}).

In the following two subsections we shall discuss two special choices
for $p(\lambda)$. In the first case all crosslinks are of unit
strength, giving rise to a point spectrum. In the second case the
strength of the crosslinks fluctuates according to 
$p(\lambda)=\exp{(-1/\lambda)/\lambda^2}$. The integral equation (\ref{Eq35})
simplifies considerably for this distribution and allows for a
solution by iteration.

\subsection{Exact solution of the integral equation for uniform crosslink
 strengths}
For crosslinks of unit strength, the integral equation (\ref{Eq35}) reduces to
Eq.\ (16) in Ref.\ \onlinecite{Bray88},
\begin{equation} \label{eq:integral}
\begin{split}\Ds 
  g^{\Omega}(\rho)  & = \hbox{} \Ds
  2c\!\,
  \exp\!\left(-\frac{i}2\,\rho^2\right) 
\left\{1 + i\, e^{-2c} \!\int_0^\infty\!\d x\,
  \rho \,\text{I}_1(i\rho x)\,
  \exp\!\left(
    \frac{i (\Omega-1)}2\,x^2 +g^{\Omega}(x) 
  \right)\right\}\\
&= 2c\!\,
  \exp\!\left(-\frac{i}2\,\rho^2\right) \left\{1 
- 2c\, e^{-2c} \!\int_0^\infty\!\d x\,
   \text{J}_1(x) \exp\!\left(
\frac{i}{2}(\Omega-1)\frac{x^2}{\rho^2} + 
  g^\Omega (\frac{x}{\rho})\right)\right\}. 
\end{split}
\end{equation}
The second equality follows from a substitution $x\to\rho x$ and from the
basic relation between the Bessel functions of the first kind $\text{J}_\nu$
and the modified Bessel functions $\text{I}_\nu$, in particular 
$\text{I}_1(x)=-i\,\text{J}_1(i\,x)$.

To get some feeling for the spectrum of eigenvalues, we first
consider the case of small $c$. We then have predominantly small
clusters, i.e.\ single particles, dimers, trimers etc. The connectivity
matrix of a dimer has eigenvalues $\lambda_1=0, \lambda_2=2$. A linear
chain of 3 particles has eigenvalues $\{0,1,3\}$, a linear chain of
four particles has eigenvalues $\{0,2,2+\sqrt{2},2-\sqrt{2}\}$ and a
star with three legs has eigenvalues $\{0,1,4\}$. These are the only
trees up to ${\cal O}(c^3)$. Hence in this order the spectrum consists
of $\delta$-functions at the above eigenvalues, with each cluster
contributing to the weight of the $\delta$-functions according to
its frequency of occurrence. Next
we show that $\delta$-functions in the spectrum correspond to Gaussian
functions $g^{\Omega}(\rho)$. The ansatz
\begin{equation}
g^{\Omega}(\rho)=2c a \exp{(-i\,z(\Omega) \rho^2/2)}
\end{equation}
where $z=z(\Omega)$ is an arbitrary function of $\Omega=\gamma+i\epsilon$ with
$\text{Im}\{z\}<0$ for $\epsilon>0$, leads to
$G(\Omega)=-1+\frac{a}{z}$.  In the limit $\text{Im}\{z\}\to 0$ each zero
$\gamma_i$ of $\text{Re}\{z(\gamma_i)\}=0$ gives rise to a $\delta$ function in the
spectrum
\begin{equation}
\label{eq:deltapeak}
D_{\text{tot}}(\gamma)=a\sum_i \frac{\delta(\gamma-\gamma_i)}
{|\partial z/\partial \gamma (\gamma_i)|}
\end{equation}

Next, we construct an approximation to the integral
equation (\ref{eq:integral}) by successive iteration. We start with
\begin{equation}
g^\Omega_0(\rho) := 2c.
\end{equation}
The first step of the iteration gives
\begin{align}
g^\Omega_1(\rho) &= 2c\exp\left(-\frac{i}{2}\rho^2\right)\left\{1 -
  e^{-2c}\int_0^\infty dx\,\text{J}_1(x)
  \exp{\left(\frac{i}{2}(\Omega-1)\frac{x^2}{\rho^2} + 
  g^\Omega_0(\frac{x}{\rho})\right)}\right\} \\
 &= 2c \exp{\left(-\frac{i}{2}\frac{\Omega}{\Omega-1}\rho^2\right)}
\end{align}
since the integral on the rhs can be calculated exactly \cite{GrRh80}.
The spectrum consists of a $\delta$-function at $\gamma=0$,
$D_1(\gamma)=\delta(\gamma)$.
The next step of the iteration gives
\begin{align}
g^\Omega_2(\rho) &= 2c\exp{\left(-\frac{i}{2}\rho^2\right)}\left\{1 -
  e^{-2c}\int_0^\infty dx\,\text{J}_1(x)
  \exp{\left(\frac{i}{2}(\Omega-1)\frac{x^2}{\rho^2} + 
  g^\Omega_1(\frac{x}{\rho})\right)}\right\}
  \label{eq:phi2a} \\
  &= 2c\exp{\left(-\frac{i}{2}\rho^2\right)}\left\{1 -
  e^{-2c}\int_0^\infty dx\,\text{J}_1(x)
  \exp{\left(\frac{i}{2}(\Omega-1)\frac{x^2}{\rho^2}\right)}
  \sum_{k=0}^{\infty}\frac{(2c)^k}{k!}
  \exp{\left(-\frac{i}{2}\frac{k\Omega}{\Omega-1}\frac{x^2}{\rho^2}\right)}\right\}
  \label{eq:phi2b}
\end{align}
by Taylor expansion of the exponential of $g^\Omega_1(\frac{x}{\rho})$.  Again,
the integrals appearing in Eq.\ \eqref{eq:phi2b} can be computed exactly,
yielding
\begin{align}
g^\Omega_2(\rho) &=
  2c \sum_{k=0}^{\infty}a^{(2)}_k 
\exp{\left(-\frac{i}{2}z^{(2)}_k\rho^2\right)}\\
a^{(2)}_k &:= e^{-2c}\frac{(2c)^k}{k!}, \quad 
z^{(2)}_k:= 
(1+\frac{1}{\Omega-1-k\frac{\Omega}{\Omega-1}}).
  \label{eq:phi2d}
\end{align}
Note that
$\sum_{k=0}^{\infty}a^{(2)}_k = 1$. In this iteration, the spectrum is
given by
\begin{equation}
\label{eq:D2}
D_2(\gamma)=\frac{1-e^{-2c}}{2c}\delta(\gamma)+\sum_{k=2}^{\infty}e^{-2c}
\frac{(2c)^{k-1}}{k(k-2)!}\delta(\gamma-k).
\end{equation}

Next, we consider a general ansatz for $g^\Omega_i$ of the form
\begin{equation}
\label{eq:phiia}
g^\Omega_i(\rho) = 
  2c \sum_{k=0}^{L}a^{(i)}_k \exp\left\{-\frac{i}{2}z^{(i)}_k\rho^2\right\},
\end{equation}
with $\sum_{k=0}^{\infty}a^{(i)}_k = 1$. $L$ is an arbitrary positive integer
and will be let tend to $\infty$ below. We insert the ansatz
(\ref{eq:phiia})
into Eq.\ (\ref{eq:integral}) and use a similar Taylor expansion as
above to obtain
\begin{equation}
g^\Omega_{i+1}(\rho)
= 2ce^{-2c}\sum_{l_0=0}^{\infty}\cdots\sum_{l_L=0}^{\infty}
  \left(\prod_{k=0}^{L}\frac{(2ca^{(i)}_k)^{l_k}}{l_k!} \right)
  \exp\left\{-\frac{i}{2}
  \left(1+\frac{1}{\Omega-1-\sum_{k=0}^{\infty}l_k z^{(i)}_k}\right)
\rho^2\right\}.
\label{eq:phii1b}
\end{equation}
When we now let $L\to\infty$, we get the expression
\begin{align}
g^\Omega_{i+1}(\rho) &= 2c \sum_{\{(l_k)\}} a^{(i+1)}_{(l_k)}
  \exp\left(-\frac{i}{2}z^{(i+1)}_{(l_k)}\rho^2\right)
\label{eq:phii1c}
\intertext{with}
a^{(i+1)}_{(l_k)} &= e^{-2c}
\prod_{k=0}^{\infty}\frac{(2ca^{(i)}_k)^{l_k}}{l_k!}
\intertext{and}
z^{(i+1)}_{(l_k)} &= 1+\frac{1}{\Omega-1-\sum_{k=0}^{\infty}l_k z^{(i)}_k}
\end{align}
We use the notation $(l_k)$ to denote a whole sequence of non-negative
integers, while $l_k$ (without parentheses) denotes the $k$-th element of the
sequence.  Out of all possible such sequences we only need those with a
\textit{finite} number of non-zero elements. This is because $a^{(i)}_k\to 0$
as $k\to\infty$, and thus
$\prod_{k=0}^{\infty}\frac{(2ca^{(i)}_k)^{l_k}}{l_k!}=0$ if there were
infinitely many non-zero elements in $(l_k)$. The set of all sequences with a
finite number of non-zero elements is denoted by $\{(l_k)\}$.
The summation in Eq.\ \eqref{eq:phii1c} thus goes over a countable set and
therefore, $g^\Omega_{i+1}(\rho)$ is of the same functional form as
$g^\Omega_i(\rho)$. It is easy to see that $\sum_{\{(l_k)\}}
a^{(i+1)}_{(l_k)}=1$ holds also for the next iteration.

Since $g^\Omega_2(\rho)$ is an expression of the form of Eq.\ 
\eqref{eq:phiia}, it follows by induction that all $g^\Omega_i(\rho)$,
$i\ge 2$, are of the same form. This observation enables us to write
down {\it fix-point} equations for the coefficients $a$ and the
exponential prefactors $z$:
\begin{equation}
\label{eq:fixpoint}
a^{(i+1)}_{(l_k)}=a^{(i)}_k \quad \text{and} \quad z^{(i+1)}_{(l_k)}=z^{(i)}_k.
\end{equation}
As shown in App.\ \ref{app:exact}, these equations can be solved if the
indices on the left and right hand sides are matched by mapping the sequence
$(l_k)$ that appears as index on the left hand side onto a simple number
$n=\sum_k l_k M^k$ with some positive integer $M$. Afterwards, $M$ is taken to
infinity. In the process, a new structure of the coefficients $a$ and $z$
emerges: each pair of coefficients $(a_i, z_i)$ falls into one of infinitely
many ``classes'' of increasing complexity. The first three classes are given
by the following expressions (the upper index denotes the class), the general
form can be found in the appendix:
\begin{xalignat}{2}
a^0_0 &= e^{-2c} & z^0_0 &= \frac{\Omega}{\Omega-1}  \\
\label{eq:class1}
a^1_n &= e^{-2c}\frac{(2ca_0^0)^n}{n!} &
  z^1_n &= \frac{\Omega-n z_0^0}{\Omega-1-n z_0^0} \\
\label{eq:class2}
a^2_{(l_k)} &= e^{-2c}\prod_{k=0}^\infty\frac{(2ca^1_k)^{l_k}}{l_k!} &
  z^2_{(l_k)} &= \frac{\Omega-\sum_{k=0}^\infty l_kz^1_k}
                      {\Omega-1-\sum_{k=0}^\infty l_kz^1_k}.
\end{xalignat}
Note that the expression for a higher class automatically contains all of the
lower classes as well if the lower-class expressions are recursively inserted,
e.g.\ $a^2_{1,0,0,\dots}=e^{-2c}\frac{(2ca^1_1)^1}{1!}=a^1_1$.  This remains
true in the general case.  For higher classes, the indices become more
complicated, e.g.\ for class 3 it is necessary to use $(l_{(k_i)})$ as index
on the left hand sides. As a shorthand, however, it is convenient to use the
notation $(l_k)$ or just $k$ even for the higher classes. It is then
understood that $k$ itself may stand for a more complicated object like a
nested sequence. See App.\ \ref{app:exact} for details.

We mention the result that $s^m$, the sum over all $a$ from
classes 0 to $m$, is given by
\begin{align}
s^m &:= \sum_{\{(l_k)\}}a^m_{(l_k)} 
 = e^{-2c}\prod_{k} e^{2ca^{m-1}_k}
 = \exp\left\{-2c(1-s^{m-1})\right\},\quad\text{and}\\
s^0 &= e^{-2c}
\end{align}
As long as $c<\minifrac{1}{2}$, the corresponding fixpoint equation, $s =
e^{-2c(1-s)}$, has a stable fixpoint at $s=1$, which implies
$\lim_{m\to\infty}s^m=1$, as it should be.  The quantity $1-s^m$ is therefore
a measure for the quality of an approximation that only goes up to class
$m$. We can conclude that for small $c$ only a few classes are sufficient
whereas for $c$ close to $\minifrac 12$ considerably more are needed. For
$c>\minifrac 12$, the fixpoint becomes unstable, indicating that the iteration
does no longer converge to the full solution of the intergral equation due to
the appearance of the infinite cluster.

\subsubsection*{Implications for the density of states}

Making use of the solution just constructed, the resolvent can be written as
\begin{equation}
G(\Omega) = -1 + \lim_{m\to\infty}\sum_{k} \frac{a^m_k}{z^m_k}.
\end{equation}
Here, inclusion of $a$'s and $z$'s from classes lower than $m$ in $a^m_k$ and
$z^m_k$ has been implied as explained above.  Analogous to Eq.\
\eqref{eq:deltapeak}, this results in the exact density of states
\begin{equation}
D_{\text{tot}}(\gamma) = \lim_{m\to\infty}\sum_{k} a^m_k \sum_i
  \frac{\delta(\gamma-\gamma^m_{ki})}
       {|\partial z^m_k/\partial\gamma(\gamma^m_{ki})|},
\end{equation}
that is, a sum of $\delta$-peaks located at the roots $\gamma^m_{ki}$ of the
respective $z^m_{k}(\gamma)$ with weight factors $a^m_k\bigg|\frac{\partial
  z^m_k}{\partial\gamma}(\gamma^m_{ki})\bigg|^{-1}$. It can be proved with
Cauchy's integration theorem applied to $(z^m_{(l_k)})^{-1}$ and Eq.\ 
\eqref{eq:class2} or the more general expression from App.\ \ref{app:exact}
that $\sum_i\bigg|\frac{\partial
  z^m_k}{\partial\gamma}(\gamma^m_{ki})\bigg|^{-1}=1$ holds for every $z^m_k$.
This property guarantees that the total weight of all peaks in the spectrum is
1 (recall that the sum of all $a$'s is also 1).  There is no continuous part
of the spectrum, but this would change for $c>\minifrac{1}{2}$ due to the
appearance of an infinite cluster.

It is impossible to find the roots of all $z^m$ but classes 0 and 1 can be
solved exactly. We deduce from Eq.\ \eqref{eq:class1} that the roots of
$z^1_n$ are located at $\gamma_{n,1} = 0$ and $\gamma_{n,2} = n+1$. The weight
factors are easily computed as $\frac 1{n+1}$ for the peak at 0 and
$\frac{n}{n+1}$ for the peak at $n+1$. The density of eigenvalues including
class 0 and 1 then reads
\begin{equation}
\label{eq:dichte1}
D_{\text{tot}}^1(\gamma) = \frac{e^{2ce^{-2c}}-1}{2c}\delta(\gamma) +
  \sum_{k=2}^\infty \frac{(2ce^{-2c})^{k}}{2ck(k-2)!}\delta(\gamma-k).
\end{equation}
Note that this is different from the result of the second iteration, Eq.\ 
\eqref{eq:D2}, although it contains the same peaks.

Another consequence of the exact solution of the integral equation is that the
density of states does \textit{not} show scaling behavior with respect to $c$,
i.e.\ it can not be written in the form $D_{\text{tot}}(\gamma)\sim
f(\gamma/\gamma^*(c))$ with some typical $\gamma^*(c)$.  This follows from the
fact that the positions of the peaks are given by the roots of the $z$'s which
are independent of $c$ and only the weights of the peaks depend on $c$. This
can obviously never result in an exact scaling form: if scaling were valid, a
small change of $\gamma^*$ would result in a small shift of the peak
positions, but they must stay fixed.  It will be shown below for the
fluctuating crosslink strengths that numerical solutions for the
eigenvalue density indicate that not even
an approximate scaling relation holds.  This view will furthermore be
supported by the results of the numerical diagonalization of random
matrices for different types of systems.

To conclude the discussion of the density of states for uniform crosslink
strengths, the spectrum from the iterative solution of the integral equation is
compared with results from numerical diagonalization of $\Gamma$ 
(for details see Sec.\ IV below). Fig.\ \ref{fig:compareA} shows the
numerically computed spectrum for $c=0.1$. Note that there is a peak at
$\gamma=1$, which is not present in Eq.\ \eqref{eq:dichte1}. This ``missing
peak'' can only be found in higher classes, e.g.\  in $z^2_{0,1,0,\ldots}=
\frac{\gamma(\gamma-1)(\gamma-3)}{\gamma^3-5\gamma^2+6\gamma^2-1}$. Other
roots that can easily be identified with peaks in the numerical results are at
$2\pm\sqrt{2}$ (stemming from $z^2_{1,1,0,\ldots}$) or at $\frac 52\pm
\frac{\sqrt{5}}{2}$ (stemming from $z^2_{0,2,0,\ldots}$). Fig.\ 
\ref{fig:compareB} shows a direct comparison between the same numerical
simulation and a few explicitly calculated peaks from classes up to class 3.
The agreement regarding the position of the peaks is excellent but some weight
is still missing from some of the peaks.  This weight is expected to be found
in higher classes and/or in different $z$'s which happen to have a root at the
same position.


\subsection{Numerical integration for special $p(\lambda)$}
The integral equation \eqref{Eq35} simplifies considerably for a special
choice of $p(\lambda)$, namely
\begin{equation} \label{20}
  p(\lambda)=\frac 1{\lambda^2}\exp\!\left\{-\frac 1\lambda\right\},
\end{equation}
implying $P_n=n!$. Inserting the ansatz 
$g^{\Omega}(\rho)=:f_{\Omega}(\rho^2/2)$ into Eq.\ (\ref{Eq32}) leads to
the following representation
\begin{equation} \label{Eq44}
  G(\Omega)=  -1 + \frac{i}{2c}\int_0^\infty\!\d x\, f_{\Omega}(x) ,
\end{equation}
where $f_{\Omega}(x)$ is the solution of the ordinary differential
equation (see Appendix A.2 for details)
\begin{equation} \label{Eq45}
  f_{\Omega}(x)=\,-i\, x \,f_{\Omega}''(x) + 
2c\,\exp\left\{ -2c+ i \Omega\,x+f_{\Omega}(x) \right\}
  ,\quad f_{\Omega}(0)=2c.
\end{equation}
This allows one to write down the general term in the asymptotic expansion
of $G(\Omega)$ for small $\Omega$. Close to the critical point the
lowest order moments are explicitly given by
\begin{equation} 
  M_1 = \frac{1}{4c}\left\{\ln(\frac{1}{1-2c})-2c \right\} ,
 \quad c\to\frac 12,
\end{equation}
\begin{equation} \label{29}
  M_2 = \frac{2}{15\,(1-2\,c)^3} + \frac{13}{60\,(1-2\,c)^2}
        + \O((1-2\,c)^{-1}), \quad c\to\frac 12,
\end{equation}
\begin{equation} \label{30}
  M_3 = \frac{47}{240\,(1-2\,c)^6} + \frac{16}{105\,(1-2\,c)^5}
        + \O((1-2\,c)^{-4}), \quad c\to\frac 12,
\end{equation}
and
\begin{equation} \label{31}
  M_4 = \frac{5762}{6435\,(1-2\,c)^9} + \frac{1159}{720720\,(1-2\,c)^8}
        + \O((1-2\,c)^{-7}), \quad c\to\frac 12,
\end{equation}
giving additional support to the conjecture about the Lifshitz tail
Eq.\ (\ref{lifshitz}). 

For a numerical evaluation of $G(\Omega)$ it is more convenient to
rewrite Eq.\ (\ref{Eq35}) in the following form,
\begin{equation} \label{Eq46}
\begin{split}\Ds 
  g^{\Omega}(\rho)  & = \hbox{} \Ds
  2c\,\sqrt{2i}\,\rho\,\text{K}_1\!\!\left(\sqrt{2i}\,\rho\right)
\\[3.5pt] \Ds & \Ds \hbox{} +
  4i c\,e^{-2c}\,\rho\,\text{K}_1\!\!\left(\sqrt{2i}\,\rho\right)
  \int_0^\rho\d\eta\,\text{I}_1\!\!\left(\sqrt{2i}\,\eta\right)\,
  \exp\!\left\{
    \frac{i \Omega}2\,\eta^2 +g^{\Omega}(\eta) 
  \right\} 
\\[3.5pt] \Ds & \Ds \hbox{} +
  4i c\, e^{-2c}\,\rho \, \text{I}_1\!\!\left(\sqrt{2i}\,\rho\right)
  \int_\rho^\infty\!\d\eta\,\text{K}_1\!\!\left(\sqrt{2i}\,\eta\right)\,
  \exp\!\left\{
    \frac{i \Omega}2\,\eta^2 +g^{\Omega}(\eta) 
  \right\} ,
\end{split}
\end{equation}
since in this representation, the integrands do not depend on $\rho$ and the
numerical integration thus needs to be done only once per iteration, resulting
in time and memory requirements only of the order of the number of integration
grid points. This allows for high precision computations of $g^\Omega(\rho)$,
$G(\Omega)$, and $D(\gamma)$.

Figures \ref{Fig1} and \ref{Fig2} show the results for the density of
eigenvalues from a numerical integration of Eq.\ \eqref{Eq46} using a Pad\'e
approximation in order to extrapolate $\Omega=\gamma+i\epsilon$ to
$\epsilon=0$.  There are several noteworthy points to be seen in these
figures: 

First, we expect to see Lifshitz tails for {\it all} $c$, $0<c<1/2$,
for asymptotically small $\gamma$. Precisely at the critical point
$D(\gamma)$ goes to a constant as $\gamma \to 0$. For crosslink
concentrations close to the critical one, we expect to see a crossover
between an approximately constant region at intermediate $\gamma$ to a
Lifshitz tail at very small $\gamma$. Since small values of $\gamma$
are hard to access numerically, this crossover makes it
difficult to observe the Lifshitz tail, except possibly for small $c$. 
For intermediate $c$ the data in Fig.\ \ref{Fig2} 
can be described approximately
by a straight line but with a slope different from $-\minifrac 12$. This
property will be confirmed by the results from the numerical diagonalization
presented below. 

A second remarkable point is that the density of states as seen in
Fig.\ \ref{Fig1} is clearly not suited to a scaling ansatz. There are
(at least) two different scales contained in the plot: the first is
the drop-off length $\gamma^0(c)$ which describes the scale on which
$D(\gamma)$ goes to 0 for small $\gamma$, the other is the position of
the maximum, $\gamma^{\text{max}}(c)$. While $\gamma^0$ goes to 0 for
$c\to\minifrac 12$, $\gamma^{\text{max}}$ evidently does not; these
two features together are obviously incompatible with a scaling ansatz
of the form $D(\gamma)\sim f(\gamma/\gamma^*(c))$ with some typical
$\gamma^*$. This finding is in agreement with the observation from the
exact solution for uniform crosslink strength where scaling was not
possible either. Here, however, the statement is even stronger than in
the previous case since even an approximate scaling relation is ruled
out. Note the peculiar feature that a second maximum appears in
$D(\gamma)$ for small $\gamma$ at the percolation threshold
$c=\minifrac 12$.  This is not an artifact and is confirmed by the
numerical diagonalization as shown below. It may even indicate the
presence of a third scale since the emergence of a maximum can already
be suspected in the curves for smaller $c$.

\subsection{Stress relaxation}

The characteristic features of the spectrum as discussed above, have
important consequences for the stress relaxation function. In
particular the Lifshitz tail in the spectrum gives rise to an
anomalous long time decay of the stress relaxation function in the sol
phase for all $c<c_{crit}$. The true asymptotic behavior of
$D(\gamma)\sim \exp(-\sqrt{h(c)/\gamma})$, which is proven rigorously
in Appendix A.3, implies $\chi(t)\sim \exp(-(t/\tau^*)^{\beta})$ with
$\beta=1/3$. However we are unable to estimate the timescale needed to
reach the asymptotic regime. For smaller times, the stress relaxation
function is characterized by effective exponents, just like the
spectra in Figs.\ \ref{Fig2},\ref{figErdoesLogDGamma} can be fitted to
Lifshitz tails with effective exponents that depend on crosslink
concentration $c$.

The divergence of the timescale $\tau^*(\epsilon)\sim \epsilon ^{-z}$
is determined by the function $h(c)$. The expansion of the resolvent
for small $\Omega$ suggests $z=3$. At the critical point the density
of eigenvalues is constant as $\gamma \to 0$, implying a logarithmic
divergence of the static shear viscosity and $\chi(t)\sim t^{-\Delta}$
with $\Delta=1$.

The absence of scaling in the density of states is also relevant for
the stress relaxation function. The presence of more than one
characteristic scale for the eigenvalues implies more than one
characteristic timescale for the stress relaxation function. As a
consequence, the stress relaxation function does not scale either.
This point will be discussed further below in the context of numerical
diagonalization of the connectivity matrix. Attempts to scale data for
the time dependent stress relaxation function fail, see 
Fig.\ \ref{figGtScaleDeltaTrue}.

\section{Numerical diagonalization} 
\label{sec:four}

\subsection{Numerical methods}
In this section the eigenvalue densities $D(\gamma)$ of three different types
of random networks are studied numerically: mean-field (MF) networks as well
as two- and three-dimensional simple square/cubic grids. For the first case,
crosslinks are allowed for all pairs $i,j$ of nodes while for the other
networks only crosslinks between neighboring nodes may appear. For the
finite-dimensional grids we apply periodic boundary conditions in all
directions. The size of the networks is denoted by $N$, with $N=L^d$ ($d=2,3$)
for the finite-dimensional cases. For the numerical treatment, we consider
random graphs with a fixed number $M$ of vertices, i.e.\ the crosslink
concentration is
$c=M/N$. Every crosslink has the same probability of occurrence. For the
implementation of the graphs on the computer, the LEDA library \cite{leda1999}
was used.
Network sizes up to $N=10000$ (MF), $N=3136$ ($d=2$) and $N=4096$
($d=3$) where studied. For each system size up to $10^4$ different
realizations of the disorder were considered (1000 for the largest
sizes). Different concentrations of the crosslinks between 0 and the
percolation threshold $c_{\text{crit}}$ were treated, where
$c_{\text{crit}}(\text{MF})=1/2$, $c_{\text{crit}}(d=2)=1$ and
$c_{\text{crit}}(d=3)\approx 0.7464$ \cite{stauffer1994}.

We consider the same two cases regarding the strength of the crosslinks as
above: Either all crosslinks have the same strength $\lambda=1$ or their
strengths are distributed randomly with the probability density given
in (\ref{20}).
%
Numerically, the random values for the strengths of the crosslinks are
drawn using the inversion method \cite{morgan1984}. A random number
$r$ is drawn which is uniformly distributed in $[0,1]$.  Then the
values of $\lambda:=-1/\ln r$ are distributed according to (\ref{20}).
For testing purposes also some systems were studied where the
strengths were uniformly distributed in the interval $[0.5,1.5]$. In
all cases no significant deviations of the measurable quantities for
different distributions could be observed. The main difference is that
for crosslinks of unit strength the distribution $D(\gamma)$ of the
eigenvalues is dominated by a sum of delta-peaks below the percolation
threshold while for crosslinks of continuous strength the distribution
$D(\gamma)$ is purely continuous (see below).

The numerical method works as follows: Random networks are created, with
constant or random crosslink strengths as needed. Then, for each graph the
connected components are determined \cite{aho1974}. For each connected
component the connectivity matrix is calculated, which is a real symmetric
matrix. Therefore, for determining its eigenvalues the QR algorithm and the
Householder method \cite{numrec1995} can be applied. Next, the eigenvalues are
sorted in increasing order.  Each connected component has one smallest
eigenvalue 0. Because of numerical errors usually the smallest eigenvalue is
not zero but quite small, depending on the distribution of the strengths of
the crosslinks. Consequently, the smallest eigenvalue is assigned the value
zero. Finally, the eigenvalues of all components are collected, sorted again,
and stored for further evaluation for each realization of the network.

\subsection{Results for the mean-field system}

First, we consider the density $D(\gamma)$ of non-zero eigenvalues for
the mean-field network at the percolation threshold $c=1/2$. Data for
the case $p(\lambda)=\delta(\lambda-1)$ have already been presented in
Fig.\ \ref{fig:compareA}. Here we consider the case where the
strengths of the crosslinks are distributed according Eq. (\ref{20}).
In Fig.\ \ref{figDGammaCpc} the resulting density is shown for
different system sizes together with the analytical result (obtained
from the numerical solution of Eq.\ \eqref{Eq46}). It can be seen
that the size $N=10000$ is already sufficient to find the analytical
behavior for a large range of eigenvalues. Especially the ``dip'' near
$\gamma=0.15$ is validated by the numerical data (see inset).  Because
of the finite system sizes, arbitrarily small eigenvalues cannot be
found, thus the numerics disagree from the analytical result in that
region. Later we will see that this raises several problems when
studying dynamical properties of the networks. Nevertheless, the
analytical result $\lim_{\gamma\to 0} D(\gamma)>0$ can indeed be
confirmed by extrapolating the numerical data to the infinite system
size.

The behavior of $D(\gamma)$ for different crosslink concentrations $c$ is
presented in Fig.\ \ref{figDGamma}. Once more, the numerical ($N=10000$) and
the analytical results agree very well. For small $\gamma$, the logarithm of
the density should behave as $\sim -\gamma^{-1/2}$ (Lifshitz tail).  Fig.\ 
\ref{figErdoesLogDGamma} shows the logarithm of $D(\gamma)$ in a double
logarithmic plot in complete analogy to the solution of the integral
equation (Fig. \ref{Fig2}). Presumably, the system size of $N=10000$ is
still too small in order to observe the asymptotic behavior of the
density of states for small eigenvalues.

\subsection{Results for the finite-dimensional systems}

Now we consider three-dimensional systems, which are believed to
describe real polymer networks more appropriately. The density of
eigenvalues for the case where all crosslinks have the same strength,
$p(\lambda)=\delta(\lambda-1)$, is shown in Fig.
\ref{figGridThreeGammaDeltaDistr} for $N=16^3$ and $c=0.2$. Similar to
the mean-field case, a collection of delta-peaks is obtained. Since
this kind of distribution is more difficult to analyze, we turn again
to the model where the strengths of the bonds have the broad
distribution (\ref{20}). Results for the largest system size
$N=16^3$ and different crosslink concentrations are shown in Fig.
\ref{figGridThreeDGammaP}. Below the percolation transition
$c_{\text{crit}}\approx 0.7464$ the distribution exhibits a maximum
and converges to 0 for small eigenvalues, similar to the mean-field
case. But at the transition $D(\gamma)$ diverges when the zero
eigenvalue is approached (see also inset).  Below we will show that
this behavior changes the divergence of the viscosity near the
percolation threshold.  The eigenvalue densities for the two-dimensional
network look qualitatively similar and are therefore not shown here.
The true asymptotic behavior as $\gamma\to 0$ is difficult to access, just as
in the mean-field case. 

The changes in the spectrum as compared to the mean-field case also
effect the stress relaxation, that we investigate next. First, the
viscosity given by
\begin{equation}
\overline{\eta}=(1-T_0(c))\int_0^{\infty}\frac{D(\gamma)}{\gamma}\,d\gamma
\end{equation}
is considered. Here, irrelevant prefactors have been omitted for simplicity,
cf.\ Eq.~\eqref{eq:viscosity2} for the complete expression.  In the
numerical calculation we compute
\begin{equation}
\eta=\frac{1}{N}\sum_{\gamma_i>0}\frac{1}{\gamma_i}
\end{equation}
for each realization and subsequently average over different
realizations 
of the
disorder to obtain $\overline{\eta}$. Whereas for the mean-field network
the viscosity diverges logarithmically for $c\to
c_{\text{crit}}$, for
finite-dimensional systems a divergence $\eta(c)\sim(c_{\text{crit}}-c)^{-k}$
is expected.  The reason for the different divergences is the manner
in which
$D(\gamma)$ behaves for small $\gamma$ at the percolation threshold: for the
mean-field network, $\lim_{\gamma\to 0}D(\gamma)$ is finite, but for the
finite-dimensional grids $D(\gamma)$ diverges as $\gamma\to 0$.  The exponent
$k$, which describes the viscosity near the percolation threshold, can be
determined, similar to the usual finite-size scaling relations
\cite{binder1988} for the percolation transition, from
\begin{equation}
\eta(c,L)=L^{-k/\nu} \tilde{\eta}((c-c_{\text{crit}})L^{1/\nu})\,,
\end{equation}
where $\tilde{\eta}$ is a universal function and $\nu$ is the exponent
describing the divergence of the correlation length when approaching the
percolation transition. The use of finite-size scaling enables us to
circumvent the problems which are posed by the lack of very small eigenvalues
of finite graphs.

By plotting $\eta L^{k/\nu}$ against $(c-c_{\text{crit}})L^{1/\nu}$ with
correct parameters $\nu$ and $k$ the datapoints for different system sizes for
$c\approx c_{\text{crit}}$ should collapse onto a single curve. We have taken
the values $\nu(d=2)=4/3$ and $\nu(d=3)=0.88$ from the literature
\cite{stauffer1994} and adjusted $k/\nu$. The best collapse near
$c_{\text{crit}}$ was obtained with $k(d=2)=1.19$ and $k(d=3)=0.75$.
The results are presented in Figs.\ \ref{figViscosThreeDDelta} ($d=3$) and
\ref{figViscosTwoDDelta} ($d=2$). The values we have obtained for the
different distributions of the crosslink strengths agree within the error
bars.

The value of $k$ for two dimensions agrees very well with the result
$k\sim 1.17$ found previously by Broderix et al.\cite{Broderix99},
using the high precision simulations of Gingold et al.\ 
\cite{Gingold1990}.  The result for the three-dimensional case
($k=0.75$)is slightly worse in comparison with $k\sim 0.71$
\cite{Broderix99,Gingold1990}. The
reason is that here only small system sizes up to $20^3$ could be
treated due to the fact that all eigenvalues are calculated. If one is
only interested in $k$, it is computationally less expensive to
compute the
Moore-Penrose inverse of the connectivity matrix. Thereby one might be
able to study system sizes as large as those used in 
Ref.\ \cite{Gingold1990}. For the realizations treated here, we
have checked other characteristic results concerning the percolation
transition, like the critical exponent $\sigma$, which describes the
behavior of the cluster-size distribution. The
finite-size scaling plots have a poor quality for this quantity, too,
 resulting in a rather low
precision of the exponent values. Additionally, we have observed a
systematic drift in our results: By including even smaller system
sizes, the scaling plot results in $k=0.89$ which differs even more
from the value obtained before.  Consequently, we believe that larger
system sizes are needed, to obtain a more reliable result for $k$ via
numerical diagonalization of random connectivity matrices.

Next, the behavior of the stress relaxation function (again omitting
irrelevant prefactors and using dimensionless time $\minifrac{2dt}{\zeta
  a^2}\to t$)
\begin{equation}
\chi(t)=(1-T_0(c))\int_0^{\infty}D(\gamma)\exp(-\gamma t)\,d\gamma
\end{equation}
is investigated, see Eq.\ (\ref{eq:stressrelax2}) for the complete
expression. The functions were obtained by first calculating
$D(\gamma)$ and then numerically integrating it. It would take too
much time on the computer to first calculate $\chi(t)$ for each
realization by directly summing up the contributions and then average
over the disorder.  Here, we have investigated the systems with the
continuously distributed crosslink strengths because they result in
continuous eigenvalue densities where it is easier to obtain stable
numerical data.

In Fig.\ \ref{figDTDelta} the numerical results for the mean-field network,
the $d=2$ and the $d=3$ model for the largest sizes ($c=c_{\text{crit}}$) are
shown. As mentioned before, the numerical simulations are restricted to finite
sizes of the networks and to a finite number of realizations of the disorder.
Therefore, the eigenvalue densities $D(\gamma)$ always have a smallest
eigenvalue $\gamma_{\min}$ with $D(\gamma)=0$ for $\gamma<\gamma_{\min}$.
Consequently, the long-time behavior is dominated by an exponential decrease,
$\exp(-\gamma_{\min}t)$, irrespective of the true form of $\chi(t)$.
This results in a negative curvature in the double-logarithmic plot for long
times.
Thus, in the numerical results, the asymptotic form of the relaxation function
is visible only for intermediate times (see Fig. \ref{figDTDelta}).  At
$c=c_{\text{crit}}$ a $\chi(t)\sim t^{-\Delta}$ behavior is expected.  By
fitting we obtain $\Delta=1.029(5)$ (mean field), 
$\Delta=0.830(2)$ ($d=3$), and
$\Delta=0.741(2)$ ($d=2$). The results for the mean-field case is known
exactly to be $\Delta=1$. The discrepancy comes again from the finite sizes of
the networks: Indeed, we have observed that for smaller networks a value of
the exponent is obtained which is even larger. So the result $\Delta=1$ seems
to be confirmed. The value for the three-dimensional grid is
compatible with the large range of results obtained in experiments
\cite{Winter87}. 

The behavior of $\chi(t)$ for different concentrations $c$ of the crosslinks
is shown in Figs.\ \ref{figGtErdoesP} (mean-field) and \ref{figthreeGtDgridP}
($d=3$).  
In both cases we find exponential decay for the longest times due to
finite system size. For intermediate times a stretched
exponential behavior $\chi(t)\sim \exp(-(t/\tau)^{\beta})$ is visible.
At least for finite system sizes the exponent $\beta$ seems to be
non-universal, we find values ranging from $\beta=0.5$ for small
crosslink concentrations down to $\beta=0.2$ close to the percolation
threshold. We suspect that the accessible times are too short to see
the true asymptotic behavior, which at least in mean-filed theory is
known to be a stretched exponential with exponent $\beta=1/3$,
resulting from the Lifshitz tail in the density of states. For small
times $\chi(t)$ decreases like $t^{-\Delta}$ and $\chi(0)=1$ by
definition.

Moreover, this variation of the exponent $\beta$ makes it impossible to
observe a scaling form $\chi(t)\sim t^{-\Delta} g(t/\tau)$, were $\tau$ is a
typical time scale which diverges like $\tau\sim (c_{\text{crit}}-c)^{-z}$
when approaching the percolation threshold. For the mean-field network, the
expectations from the Lifshitz tails are $z=3$ and $\Delta=1$, while $g(t)$ is
the stretched exponential function, but it was already mentioned that
there seems to be no scaling possible due to the existence of more than one
scale.  In Fig.\ \ref{figGtScaleDeltaTrue} a scaling plot of $\chi(t)$ is
shown. $\chi(t)t^{\Delta}$ is plotted against $t\times(c_{\text{crit}}-c)^z$
for mean-field networks having different concentrations $c$. We have used only
the regions below the finite-size asymptotic behavior ($\beta=1$).  It can be
seen that the quality of the collapse is rather bad, explained by the
variation of $\beta$ with $c$. One might think that near the transition
$c\approx c_{\text{crit}}$ the scaling may be better. But there the collapse
is even worse (not shown), because even larger systems are necessary to reach
the asymptotic regime for the small eigenvalues, as explained before.

For the finite-dimensional systems, the quality of the scaling-plot is
similar. Therefore, it is not possible to make a reliable estimate for
the dynamical exponent $z$ in that case.

\section{Conclusions}
\label{sec:five}

Within our model, the dynamics of a crosslinked polymer melt is determined
completely by the eigenvalue and eigenvector spectrum of the connectivity
matrix $\Gamma$. In this paper we have focused on some properties which are
determined by the eigenvalues alone (e.g.\ the stress relaxation function)
since the eigenvectors are hard to obtain.  We have used three different
methods to examine the eigenvalue spectrum: First, the construction of an
exact solution for the averaged eigenvalue density for a fixed crosslink
strength, second, a very precise numerical solution for the case of varying
crosslink strengths, and third, a numerical diagonalization of random
connectivity matrices.

The first method allowed for some exact results regarding the eigenvalue
spectrum. It could be shown that the eigenvalue spectrum consists of a very
complicated but countable set of $\delta$-peaks, some of which could be
calculated and compared with results from numerical diagonalization.
Furthermore it showed that the eigenvalue density does not show (exact)
scaling behavior.

The second model of fluctuating crosslink strengths has the advantage that
the eigenvalue spectrum becomes a continuous function instead of an
inscrutable sum of $\delta$-peaks. Additionally, it allowed for a fast
numerical integration scheme. From these numerical solutions it could be
inferred that the expected Lifshitz-tail behavior for small $\gamma$ only
seems to set in for extremely small $\gamma$, smaller than were accessible
numerically. For this reason, the stress relaxation function does not show a
stretched exponential form with exponent $\beta=\minifrac 13$ within the
accessible time window. Instead, for the times that could be reached, there
seems to be a regime where an apparent stretched exponential with a crosslink
concentration-dependent and thus non-universal $\beta$ can be observed.
Furthermore, in numerical evaluations of the eigenvalue spectrum again 
scaling could not be observed, not even approximately, since at least two,
possibly three or more, different $\gamma$-scales with different
$c$-dependence could be identified. As a consequence, the stress relaxation
function does not scale either.

The third method, numerical diagonalization, confirmed all results obtained so
far very well. In particular it showed that the stress relaxation shows
stretched-exponential behavior with a concentration-dependent exponent $\beta$
and it showed the failure of scaling of the stress relaxation function.  It
confirmed, however, the experimental findings that at the critical
concentration, the stress relaxation function decays algebraically with
exponent $\Delta$. For the mean-field model, both theory and numerics yield
the exponent $\Delta=1$.  Furthermore, numerical diagonalization allows for
going beyond the mean-field approach. Results were obtained for connectivity
matrices on two- and three-dimensional cubic lattices. Unlike the mean-field
case, the density of eigenvalues now diverges at the critical concentration as
$\gamma\to 0$.  Apart from this difference, the other results are
qualitatively very similar to the mean-field case except that the exponents
$\Delta$ and $\beta$ have different values.

\section{Acknowledgements}
We thank P. M\"uller for interesting discussions and a careful reading
of the manuscript. K.B. obtained financial support by the DFG ({\em
  Deutsche Forschungsgemeinschaft}) under grant Br 1894/1-1 , T.A. under grant
Zi209/5-1 and A.K.H. under grant Zi209/6-1.

\appendix
\section{Low frequency expansion of the resolvent}
\subsection{General $p(\lambda)$}

The low frequency expansion is derived from an alternative form of the
integral equation (\ref{Eq35}). We start from Eq.\ (\ref{Eq25}) and recall an
integral representation of the $n$-dimensional Laplacian (see also
Eqs.\ (3.47-51)) in Ref.\ \onlinecite{Broderix2000}
\begin{equation}
\int
\frac{d\hat{y}}{(2\pi\Omega)^{n/2}}
\exp{\left\{-\frac{(\hat{x}-\hat{y})^2}{2\Omega}\right\}}f(|\hat{y}|)= 
\left.\exp{\left\{\frac{\Omega}{2}\left(\frac{d^2}{d\rho^2}+\frac{n-1}{\rho}
  \frac{d}{d\rho}\right)\right\}}
f(\rho)\right|_{\rho=|\hat{x}|}.
\end{equation}
We use this representation in the numerator of Eq.\ (\ref{Eq25}) and
take the limit $n\to 0$.  To evaluate the denominator of
Eq.\ (\ref{Eq25}) we observe that
\begin{equation}
\lim_{n\to 0} \int d\hat{x} f_n(|\hat{x}|)=f_0(0)+{\cal O}(n).
\end{equation}
Both steps taken together lead to the following self-consistent
equation for $g^{\Omega}(\rho)$ 
\begin{equation} \label{Eq34}
  g^{\Omega}(\rho)=2c\,e^{-2c}\,\int_0^\infty\!\d\lambda\,p(\lambda)\,
  \exp\!\left\{
    \frac 1{2 i \lambda}
    \left(\frac{\partial^2}{\partial\rho^2}
    -\frac{\partial}{\rho\,\partial\rho}\right)
  \right\}
  \:\exp\!\left\{
    \frac{i\Omega}2\rho^2+g^{\Omega}(\rho)
  \right\},
\end{equation}
which is of course equivalent to the integral equation (\ref{Eq35}),
but much better suited for a low frequency expansion.

To that end we rescale variables according to $x=\sqrt{\Omega}\rho$
and $\Psi^{\Omega}(x)=g^{\Omega}(x/\sqrt{\Omega})$. The self consistent
equation then reads
\begin{equation} \label{Eq54}
  \Psi^{\Omega}(x)=2c\,e^{-2c}\,\int_0^\infty\!\d\lambda\,p(\lambda)\,
  \exp\!\left\{
    \frac {\Omega}{2 i \lambda}
    \left(\frac{\partial^2}{\partial x^2}
    -\frac{\partial}{x\,\partial x}\right)
  \right\}
  \:\exp\!\left\{
    \frac{i}2 x^2+\Psi^{\Omega}(x)
  \right\}.
\end{equation}
We look for a solution in terms of a power series in $\Omega$
\begin{equation}
\Psi^{\Omega}(x)=\sum_{j=0}^{\infty}(\Omega)^j \Psi_j(x).
\end{equation}
The resolvent can then be expressed in terms of $\Psi_j(x)$ as follows
\begin{equation}
G(\Omega)=- P_1+\frac{i}{2c\Omega}\sum_{j=0}^\infty \Omega^j
\int_{0}^{\infty}dx x \Psi_j(x)
\end{equation}
with $ P_n=\int_0^\infty \d\lambda \lambda^{-n}\,p(\lambda)$
as defined after Eq.\ (\ref{14}).
The lowest order term obeys the equation
\begin{equation}
\label{Lambert1}
\Psi_0(x)= 2c \exp{(-2c)} \exp{\left(\frac{i}{2}x^2+\Psi_0(x)\right)},
\end{equation}
which is solved by 
\begin{equation}
\Psi_0(x)= - W\left(-2c\exp{(-2c)}\exp{(ix^2/2)}\right).
\end{equation}
Here $W$ denotes the principal branch of Lambert's $W$-function, defined
as the solution of 
\begin{equation}
W(x)\exp{(W(x))}=x.
\end{equation}
From Eq.\ (\ref{Lambert1}) one derives the following property of the
lowest order solution
\begin{equation} 
\Psi_0'(x)=\frac{ix\Psi_0(x)}{1-\Psi_0(x)}
\end{equation}
which allows for an exact computation of the integral
\begin{equation}
\label{Lambert2}
i\int_{0}^{\infty}dx x \Psi_0(x)=-\frac{1}{2}\int_{0}^{\infty}dx
\frac{d}{dx}\left (1-\Psi_0(x)\right)^2 =2c(c-1).
\end{equation}

The next two terms are given by
\begin{eqnarray}
\Psi_1(x) & = & \frac{1}{1-\Psi_0(x)}\frac{P_1}{2i}
\left(\frac{d^2}{dx^2}
    -\frac{d}{x\, dx}\right) \Psi_0(x) \\
\Psi_2(x) & = & \frac{-1}{1-\Psi_0(x)}\left(\frac{d^2}{dx^2}
    -\frac{d}{x\,dx}\right)\left(\frac{P_2}{8}+\frac{P_1^2}{1-\Psi_0(x)}\right)
\left(\frac{d^2}{dx^2} -\frac{d}{x\,dx}\right) \Psi_0(x).
\end{eqnarray}
The integrals $\int dx x \Psi_j(x)$ can be performed similar to
Eq.\ (\ref{Lambert2}), using the properties of Lambert's $W$-function. The
computations, however, become increasingly tedious, so that higher
order terms have only been computed for the special distribution
$p(\lambda)$, see below. 
 
\subsection{Special $p(\lambda)$}

We start from Eq.\ (\ref{Eq34}) and introduce the abbreviation 
$D_{\rho}:=\frac{d^2}{d\rho^2} -\frac{d}{\rho\,d\rho}$. For the
special choice $p(\lambda)=\frac{1}{\lambda^2}\exp{(-1/\lambda)}$, one
can perform the average over $p(\lambda)$ analytically
\begin{equation}
\int_{0}^{\infty} \frac{d\lambda}{\lambda^2}
\exp{\left\{-\frac{(1+iD_{\rho}/2)}{\lambda}\right\}} 
\exp\!\left\{
    \frac{i\Omega}2\rho^2+g^{\Omega}(\rho)
  \right\} =\left(1+\frac{iD_{\rho}}{2}\right)^{-1}
\exp\!\left\{
    \frac{i\Omega}2\rho^2+g^{\Omega}(\rho)
  \right\}.
\end{equation}
The resulting differential equation simplifies, if we introduce the
function $f_{\Omega}(\rho^2/2):=g^{\Omega}(\rho)$ 
\begin{equation}
f_{\Omega}(\rho^2/2)+i\rho^2/2f''_{\Omega}(\rho^2/2)=
2c\exp{(-2c)}\exp{(i\Omega\rho^2/2+f_{\Omega}(\rho^2/2))}.
\end{equation}
Introducing the new variable $x=\rho^2/2$ leads to the differential
Eq.\ (\ref{Eq45}) quoted in the main part of the paper. For the low
frequency expansion it is convenient to introduce yet another
variable, $y=\Omega x$, in terms of which the differential
equation  for $h_{\Omega}(\Omega x):=f_{\Omega}(x)$reads
\begin{equation}
h_{\Omega}(y)-i y \Omega h''_{\Omega}(y)=2c\exp{(-2c)}\exp{(iy+h_{\Omega}(y))}.
\end{equation}
The ansatz $h_{\Omega}(y)=\sum_{j=0}^{\infty}(\Omega)^j h_j(y)$ then
yields
\begin{equation}
\label{Lambert3}
h_n(y)-iyh''_{n-1}(y)=h_0(y)\frac{1}{n!}\frac{d^n}{d\Omega^n}
\exp{\left(\sum_{j=1}^{\infty}(\Omega)^j h_j(y)\right)}_{\Omega=0}.
\end{equation}
The left hand side is linear in $h_n(y)$, so that Eq.\ (\ref{Lambert3})
is easily iterated.

\subsection{Proof of the existence of a Lifshitz tail in $D(\gamma)$}

The aim of this appendix is to prove that the density of eigenvalues
$D(\gamma)$ shows a Lifshitz-tail behavior for $\gamma\to 0$ and $c<1/2$. For
the proof, it is convenient to make use of the eigenvalue distribution
function, $F(\gamma):=\int_{-\infty}^\gamma\drm\gamma'\,D(\gamma')$.  This can
be done without loss of generality because if $F(\gamma)$ has a Lifshitz tail,
so does $D(\gamma)$. It will be shown that $F(\gamma)$ lies between two bounds
which, taken together, assert the Lifshitz behavior.

For a given realization of a system with $N$ vertices (or
polymers), the corresponding $F_N(\gamma)$ can be written, using a
decomposition into the $K$ clusters of the realization,
\begin{equation}
F_N(\gamma) = \frac{1}{N}\sum_{k=1}^K\Tr \Theta(\gamma-\Gamma_k)
= \frac{K}{N} + \frac{1}{N}\sum_{k=1}^K\Tr[(1-E_0^k)\Theta(\gamma-\Gamma_k)]
\end{equation}
where $\Gamma_k$ is the connectivity matrix of the $k$-th cluster and $E_0^k$
is the projector on the nullspace of $\Gamma_k$. In the macroscopic limit,
$N\to\infty$, this yields
\begin{equation}
\label{eq:Fself}
F(\gamma) = (1-c)\Theta(\gamma) + \sum_{n=1}^\infty \tau_n
  \left\langle\Tr[(1-E_0^n)\Theta(\gamma-\Gamma({\mathcal{T}}_n))]\right\rangle
\end{equation}
due to self-averaging. The bracket $\langle\cdots\rangle$ means averaging over
the set of all numbered trees $\{{\mathcal{T}}_n\}$ of size $n$ of which there
are $n^{n-2}$.  $\Gamma({\mathcal{T}}_n)$ denotes the connectivity matrix
corresponding to the tree ${\mathcal{T}}_n$. The average number of trees of size
$n$ per vertex is denoted by $\tau_n$ and is given by \cite{ErRe60}
\begin{equation}
\tau_n=\frac{n^{n-2}}{2cn!}(2ce^{-2c})^n
=\frac{1}{2c\sqrt{2\pi}}n^{-5/2}e^{-nh(c)-f(n)/n}
\end{equation}
according to Stirling's formula with $h(c)=2c-1-\ln(2c)$ and some function
$f(n)$ with $0<f(n)<1$.

The smallest non-zero eigenvalue of $\Gamma({\mathcal{T}}_n)$ is certainly
greater than or equal to the smallest non-zero eigenvalue of the linear
cluster with $n$ vertices, which is proportional to $n^{-2}$, i.e.\
\begin{equation}
\Gamma({\mathcal{T}}_n)\ge\frac{\alpha}{n^2}
\end{equation}
(except for the zero eigenvalue) with some $\alpha$ independent of $n$. This
results in
\begin{equation}
\Tr[(1-E_0^n)\Theta(\gamma-\Gamma({\mathcal{T}}_n))]\le
   (n-1)\Theta(\gamma-\alpha/n^2)
\end{equation}
or
\begin{equation}
F(\gamma)\le 1-c +
\sum_{n\ge\sqrt{\alpha/\gamma}} (n-1)\tau_n,\quad\text{for }\gamma>0.
\end{equation}
For $\gamma\to 0$, the sum can be approximated by an integral,
\begin{align}
F(\gamma) &\le 1-c +
\frac{1}{2c\sqrt{2\pi}}\int_{\sqrt{\alpha/\gamma}}^\infty n^{-3/2}e^{-nh(c)}\\
&\approx 1-c + \frac{1}{2ch(c)\sqrt{2\pi}}\left(\frac{\gamma}{\alpha}\right)^{3/4}
\e{-h(c)\left(\frac{\alpha}{\gamma}\right)^{1/2}}.
\end{align}
This is the lower bound for $F(\gamma)$.

For the upper bound, Eq.~\eqref{eq:Fself} will be used again. Explicitly, one
has for $\gamma>0$
\begin{align}
F(\gamma) &= 1-c +
\frac{1}{2c}\sum_{n=1}^\infty \frac{1}{n!}(2ce^{-2c})^n\sum_{\{{\mathcal{T}}_n\}}
\Tr[(1-E_0^n)\Theta(\gamma-\Gamma({\mathcal{T}}_n))] \\
&\ge 1-c + 
\frac{1}{2c}\sum_{n=1}^\infty \frac{1}{n!}(2ce^{-2c})^n\sum_{\{\mathcal{L}_n\}}
\Tr[(1-E_0^n)\Theta(\gamma-\Gamma(\mathcal{L}_n))]
\end{align}
where the inner sum has been restricted on the set of \textit{linear} numbered
trees $\{\mathcal{L}_n\}$. There are $n!/2$ such linear trees, such that
\begin{equation}
F(\gamma) = 1-c +
\frac{1}{4c}\sum_{n=2}^\infty e^{-n(h(c)+1)}
\Tr[(1-E_0^n)\Theta(\gamma-\Gamma(\mathcal{L}_n))].
\end{equation}
Next, the trace, which is a sum of non-negative terms, is estimated by just
one of the terms. In particular,
$\Tr[(1-E_0^n)\Theta(\gamma-\Gamma(\mathcal{L}_n))] \ge
\Theta(\gamma-\alpha/n^2)$, corresponding to the smallest eigenvalue of
$\mathcal{L}_n$. This finally gives
\begin{align}
F(\gamma) &\ge 1-c + \frac{1}{4c}\sum_{n\ge\sqrt{\alpha/\gamma}}
  e^{-n(h(c)+1)} \\
&\approx 1-c + \frac{1}{2c(1+h(c))}
  \e{-\left(\frac{\alpha}{\gamma}\right)^{1/2}(1+h(c))}
\end{align}
for the lower bound.

The upper and the lower bound together imply
\begin{equation}
\lim_{\gamma\to 0}\frac{\ln|\ln(F(\gamma)-1+c)|}{|\ln\gamma|} = \frac{1}{2}
\end{equation}
or, even stronger,
\begin{equation}
\sqrt{\alpha}h(c) \le -\lim_{\gamma\to 0}\gamma^{1/2}\ln(F(\gamma)-1+c)
                  \le \sqrt{\alpha}(h(c)+1),
\end{equation}
which is the sought-for Lifshitz tail behavior.

\section{Details of the exact solution of the integral equation}
\label{app:exact}
\subsection{Solution of the integral equation}
It is not obvious how to solve the fix-point equations \eqref{eq:fixpoint},
because the coefficients of the $i$-th iteration are labeled by an index, and
a subsequent iteration gives rise to coefficients which are labelled by a
sequence $(l_k)$.  We therefore try to map the sequence $(l_k)$ that appears
as index onto a number by writing $n:=\sum_{k=0}^{\infty}l_k M^k$ with some
$M\in\mathbb{N}$. For this to be a one-to-one map, we need to restrict all
$l_k$ to be $<M$. This restriction will be removed later when we let
$M\to\infty$. The sequence $(l_k)$ can be reconstructed from $n$ by writing
$n$ in the number system of base $M$. Let this be indicated by $l_k =
(n)^M_k$.

The fix-point equations can now be written down as:
\begin{align}
\label{eq:fixa}
a_n &= e^{-2c}
  \prod_{k=0}^{\infty}\frac{(2ca_k)^{(n)^M_k}}{(n)^M_k!} \\
\label{eq:fixz}
z_n &= 1+\frac{1}{\Omega-1-\sum_{k=0}^{\infty}(n)^M_k z_k}.
\end{align}
The equations for $a_n$ can be solved independently from those for $z_n$. We
start with $a_n$. Successively solving the system of equations \eqref{eq:fixa}
by inspection gives
\begin{align}
a_0 &= e^{-2c} \\
a_n &= e^{-2c}\frac{(2ca_0)^n}{n!}\quad\text{ for } 1 \le n < M \\
a_n &= e^{-2c}\prod_{k=0}^{M-1}\frac{(2ca_k)^{(n)^M_k}}{(n)^M_k!}\quad
       \text{ for } M \le n < M^M \\
a_n &= e^{-2c}\prod_{k_0=0}^{M-1}\cdots\prod_{k_{M-1}=0}^{M-1}
  \frac{(2ca_{k_0+Mk_1+\cdots+M^{M-1}k_{M-1}})^{(n)^M_{k_0+Mk_1\cdots}}}
  {(n)^M_{k_0+Mk_1\cdots}!}\quad\text{ for } M^M \le n < M^{M^M}
\end{align}
and so on. The coefficient $a_0$ is obviously independent of all other $a_n$.
This property will be called ``class~0''. $a_1,\ldots,a_{M-1}$ only depend on
$a_0$: this will be termed ``class~1''. Analogously, $a_M,\ldots,a_{M^M-1}$
are in class~2 as they only depend on $a$'s from classes 0 and 1.

Now we can let $M$ tend to infinity. Classes 0 and 1 are simple (the upper
index now denotes the class):
\begin{align}
a^0_0 &= e^{-2c}\\
\label{eq:a1}
a^1_n &= e^{-2c}\frac{(2ce^{-2c})^n}{n!},\quad n\ge 1
\end{align}
If we drop the constraint $n\ge 1$, Eq.\  \eqref{eq:a1} automatically contains
class 0. 

For the higher classes, as we are now considering $M\to\infty$, indexing
via a number $n$ is no longer possible. Instead, for class 2, we have to
revert to using a finite sequence as index. For class 3, even this is not
sufficient and a nested sequence $(l_{(k_i)})$ is needed:
\begin{align}
\label{eq:a2}
a^2_{(l_k)} &= e^{-2c}\prod_{k=0}^{\infty}\frac{(2ca^1_k)^{l_k}}{l_k!},
  \quad \text{length of }(l_k) > 1, \\
\label{eq:a3}
a^3_{(l_{(k_i)})} &= e^{-2c}\prod_{\{(k_i)\}}
\frac{(2ca^2_{(k_i)})^{l_{(k_i)}}}   {l_{(k_i)}!}.
\end{align}
If the constraint (length of $(l_k) > 1$) is dropped and if the explicit
expressions for the $a$ from the lower classes are recursively inserted, all
classes up to class 2 are contained in one formula, Eq.\ (\ref{eq:a2}).  An
analogous statement holds for Eq.\ \eqref{eq:a3}.

In general, for class $m$, the index will be of the form
$(l_{(k_{\ddots_{(r_i)}})})$ with $m$ nesting levels. The general result is
thus
\begin{align}
a^m_{(l_{(k_{\ddots_{(r_i)}})})} &= 
  e^{-2c}\prod_{\big\{(k_{\ddots_{(r_i)}})\big\}}
  \frac{\bigg(2ca^{m-1}_{(k_{\ddots_{(r_i)}})}\bigg)
  ^{l_{(k_{\ddots_{(r_i)}})}}}
  {l_{(k_{\ddots_{(r_i)}})}!}.
\end{align}

With the same reasoning as above we can calculate the $z_n$. We find the same
classes, and the results are:
\begin{align}
z^0_0 &= \frac{\Omega}{\Omega-1} \\
\label{eq:z1}
z^1_n &= \frac{\Omega-nz^0_0}{\Omega-1-nz^0_0} \\
z^2_{(l_k)} &= \frac{\Omega - \sum_{k=0}^{\infty}l_k z^1_k}
                  {\Omega - 1 - \sum_{k=0}^{\infty}l_k z^1_k}\\
&\vdots\nonumber\\
z^m_{(l_{(k_{\ddots_{(r_i)}})})} &= \frac{\Omega - 
  \sum_{\big\{(k_{\ddots_{(r_i)}})\big\}}
  l_{(k_{\ddots_{(r_i)}})}z^{m-1}_{(k_{\ddots_{(r_i)}})}}
  {\Omega - 1 -
  \sum_{\big\{(k_{\ddots_{(r_i)}})\big\}}
  l_{(k_{\ddots_{(r_i)}})}z^{m-1}_{(k_{\ddots_{(r_i)}})}}.
\end{align}

\subsection{Properties of the solution}

If one asks for the total weight of a particular peak at, say, some $\gamma_0$
(up to class $m$), one has to find all finite solutions $(l_k)$ of the
diophantic equation
\begin{equation}
\label{eq:diophant}
\gamma_0 - \sum_{k=0}^\infty l_k z^{m-1}_k(\gamma_0) = 0.
\end{equation}
This is possible in some special cases, e.g.\ for $\gamma_0=1$ in class 2.
Since $z^1_n(1)=1$, it follows that Eq.\ \eqref{eq:diophant} is satisfied if
and only if exactly one entry of $(l_k)$ equals 1 whereas all the others are
0. Adding up all of the weights yields $e^{-2c}\left(\left(2ce^{-2c}-1\right)
  e^{2ce^{-2c}}+1\right)$ as the total weight of $\delta(\gamma-1)$ from class
2.

The $z^m_{(l_k)}$ have several noteworthy properties most of which are easy to
prove by induction over $m$ and are therefore listed below without proof.
\begin{enumerate}
\item $z^m_{(l_k)}$ is a rational function of $\gamma$ with integer
coefficients.
\item The degree of the numerator is the same as that of the denominator.
\item The coefficient of the highest power is 1 in both numerator and
  denominator.
\item $z^m_{(l_k)}$ is a strictly monotonically decreasing function (except at
its poles).
\item All roots and poles of $z^m_{(l_k)}$ are located on the non-negative
  real axis.
\item $z^m_{(l_k)}$ has exactly as many poles as roots. Roots and poles
  alternate, starting with a root at 0.
\item There are exactly one more roots of $z^m_{(l_k)}$ than there are poles
in $\sum_{k=0}^\infty l_k z^{m-1}_k$.
\item The sum $\sum_{i}\left|\frac{\partial
      z^m_{(l_k)}}{\partial\gamma}(\gamma^m_{(l_k)i})\right|^{-1}$ over all
  roots $\gamma^m_{(l_k)i}$ of $z^m_{(l_k)}$ equals 1. As stated in the text,
  this can be proved using Cauchy's integration theorem.
\end{enumerate}

Consider now some $z^m_{(l_k)}$ and choose $(l_k)$ such that only the $n$-th
entry is nonzero. Then we have
\begin{equation}
z^m_{0,\ldots,0,l_n,0,\ldots} = 
  \frac{\gamma - l_n z^{m-1}_n}{\gamma - 1- l_n z^{m-1}_n}.
\end{equation}
Between two of its poles (see the list of properties above), $z^{m-1}_n$ is a
continuous function that maps bijectively onto the real numbers; therefore
there exists a $\gamma^m_{l_n}$ in this interval such that
$z^m_{(l_k)}(\gamma^m_{l_n})=0$. Moreover, when $l_n\to\infty$, the
$\gamma^m_{l_n}$ converge to the root $\gamma^{m-1}_{n,i}$ of $z^{m-1}_n$ in
this interval. Since $z^{m-1}_n$ is monotonously decreasing, $\gamma^m_{l_n} <
\gamma^{m-1}_{n,i}$. This implies that for every peak in the spectrum there
are infinitely many other peaks to the left of it in any arbitrarily small
interval around this peak. This also applies recursively for each of these
satellite peaks! Only the peak at 0 is different: as stated in the list above,
all roots of $z^m_{(l_k)}$ are $\ge 0$ and thus there are no satellite peaks
of $\delta(\gamma)$.

\newcommand{\captionCompareA}
{Numerical simulation of the density of states for $c=0.1$.}

\newcommand{\captionCompareB}
{Comparison between the simulation (solid lines) and some selected peaks
  calculated from the exact solution (dashed lines) for $c=0.1$. The
  analytical peaks have been slightly shifted to the right for better
  comparison, otherwise they would be indistinguishable from the numerical
  peaks.}

\newcommand{\captionFigOne}
{  Density $D(\gamma)$ of non-zero eigenvalues as a function of
  $\gamma$ for $p(\lambda)$ given by (\ref{20}) for
  several concentrations $c$. For $c=0$, the density is
  given by $D(\gamma)=\minifrac 12 p(\minifrac\gamma 2)$.}

\newcommand{\captionFigTwo}
{  Double logarithm of the density $D(\gamma)$ of non-zero eigenvalues as a
  function of $\ln\gamma$ for several concentrations $c$.}

\newcommand{\captionDGammaCpc} {Density $D(\gamma)$ of non-zero eigenvalues
  for the mean-field network at the percolation threshold $c=0.5$
from numerical diagonalization. The solid line is
  the analytical result, which is hardly distinguishable from the result for
  $N=10000$.  The inset magnifies the region $\gamma \in [0,0.4]$, where the
  numerical results for the largest system size $N=10000$ are shown by
  circles.}

\newcommand{\captionDGamma} {Density $D(\gamma)$ of non-zero eigenvalues for
  the mean-field network for different concentrations $c$. The lines are the
  analytical results while the numerical data is shown by the symbols.}
\newcommand{\captionErdoesLogDGamma}
{Double logarithmic plot of $-\ln (D(\gamma))$ for different
  concentrations $c$ of the mean-field network.
 The line shows a function $-\ln(\gamma)/2+const$ (Lifshitz tail), which
 is the behavior predicted by theory.}

\newcommand{\captionGridThreeGammaDeltaDistr} {Density $D(\gamma)$ of
  non-zero 
eigenvalues for the cubic network with all bonds having the same strength
  $\lambda=1$ ($c=0.2,N=16^3$). Similar to the case of the mean-field network,
  a sum of delta-peaks with strongly varying heights is obtained.}

\newcommand{\captionGridThreeDGammaP} {Density $D(\gamma)$ of non-zero
  eigenvalues for
  the cubic network with $p(\lambda)$ given by (\ref{20})
  for different concentrations $c$. The inset shows the finite-size dependence
  at the percolation threshold for small eigenvalues.}

\newcommand{\captionGridThreeLogDGamma} {Double logarithmic plot of $-\ln
  (D(\gamma))$ of the cubic network (continuously distributed strengths) for
  different concentrations $c$ .  The line shows a function $\sim
  \gamma^{0.5}$.}

\newcommand{\captionViscosThreeDDelta} {Finite-size scaling plot of the
  viscosity $\eta(c,L)$ for the three-dimensional grid. A scaling behavior of
  $\eta(c,L)=L^{-k/\nu} \tilde{\eta}((c-c_{\text{crit}})L^{1/\nu})$ is
  assumed. Using $\nu=0.88$ and $k=0.75$ the points for $L=10,13,16,20$
  collapse onto one curve near the critical concentration.}

\newcommand{\captionViscosTwoDDelta} {Finite-size scaling plot of the
  viscosity $\eta(c,L)$ for the two-dimensional grid. A scaling behavior of
  $\eta(c,L)=L^{-k/\nu} \tilde{\eta}((c-c_{\text{crit}})L^{1/\nu})$ is
  assumed. Using $\nu=4/3$ and $k=1.19$ the points for
  $L=10,14,20,28,40,56$ collapse onto one curve near the critical
  concentration. Since finite systems are treated, the maximum of
  $\eta(c)$ is below the critical concentration $c_{\text{crit}}=1$
of the infinite lattice.}  

\newcommand{\captionGTDelta} {Stress relaxation function $\chi(t)$ at the the
  critical concentration $c=c_{\text{crit}}$ for the three types of models
  considered here, with continuously distributed strengths of the crosslinks
  in all three cases.  Shown are the results for the largest sizes which could
  be treated with sufficient accuracy.  For the part of the long-time behavior
  which is accessible to the numerical simulations, a $\chi(t)\sim
  t^{-\Delta}$ behavior is visible. From fitting we obtain $\Delta=1.029$
  (mean field), $\Delta=0.830(2)$ (d=3) and $\Delta=0.741(2)$ (d=2).  }

\newcommand{\captionGtErdoesP} {Rescaled stress relaxation function
  $-\ln(\chi(t)t^{\Delta})$ as a function of the time for the mean-field
  network ($\Delta=1.029$) with different concentrations $c$ of the
  crosslinks. The straight lines correspond to stretched exponentials with
  exponent $\beta=0.332$ and $\beta=1$.  }

\newcommand{\captionGtThreeDgridP} {Rescaled stress relaxation function
  $-\ln(\chi(t)t^{\Delta})$ as a function of the time for the
  three-dimensional network ($\Delta=0.830$) with different concentrations $c$
  of the crosslinks. The straight lines correspond to stretched exponentials
  with exponent $\beta=0.386$ and $\beta=1$.  }

\newcommand{\captionGtTwoDgridP} {Rescaled stress relaxation function
  $-\ln(\chi(t)t^{\Delta})$ as a function of the time for the two-dimensional
  network ($\Delta=0.741$) with different concentrations $c$ of the
  crosslinks. The straight lines correspond to stretched exponentials with
  exponent $\beta=0.362$ and $\beta=1$.  }

\newcommand{\captionGtScaleDeltaTrue} {Scaling plot for the stress relaxation
  $\chi(t)t^{\Delta}$  as a function of
  $t(c_{\text{crit}}-c)^z$ for the mean-field network ($N=10^4$, randomly
  distributed strengths of crosslinks) with the values $\Delta=1,z=3$.}

\begin{figure}
\begin{center}
\myscaleboxb{\includegraphics{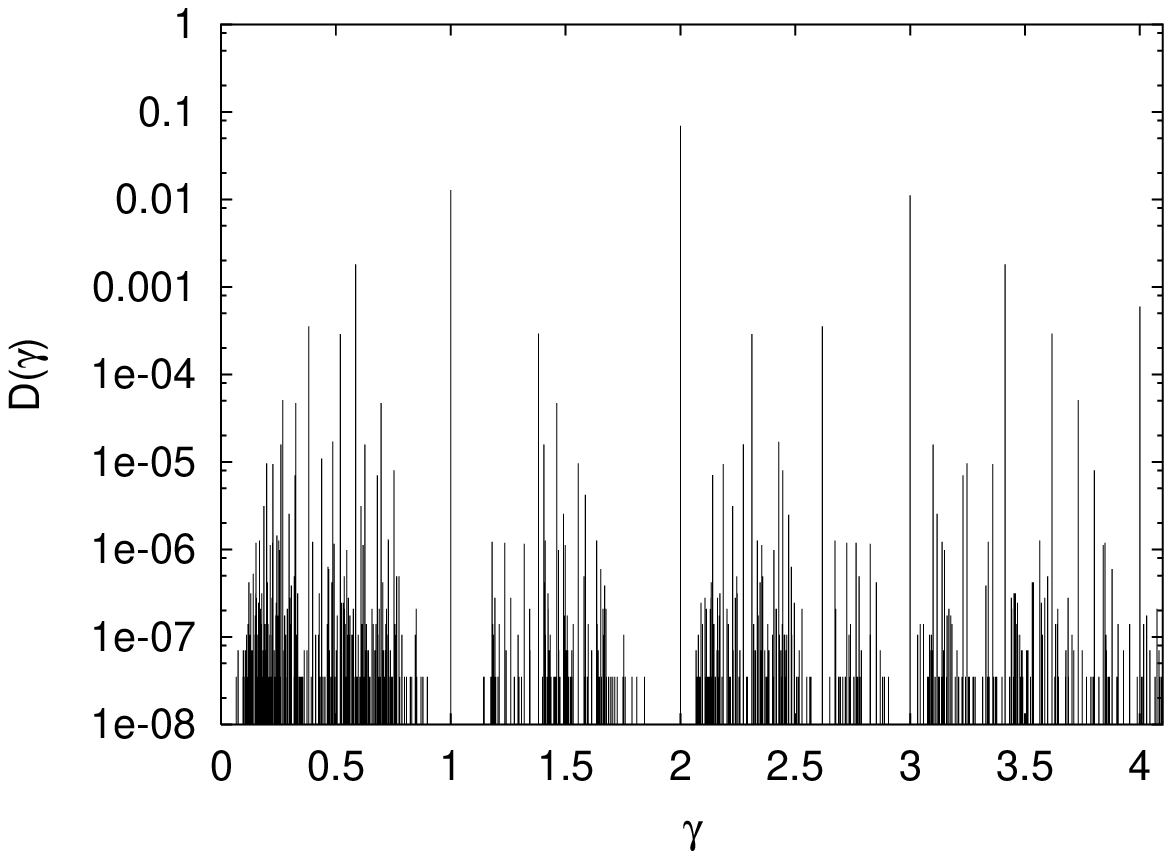}}
\end{center}
\caption{\captionCompareA}
\label{fig:compareA}
\end{figure}

\begin{figure}
\begin{center}
\myscaleboxb{\includegraphics{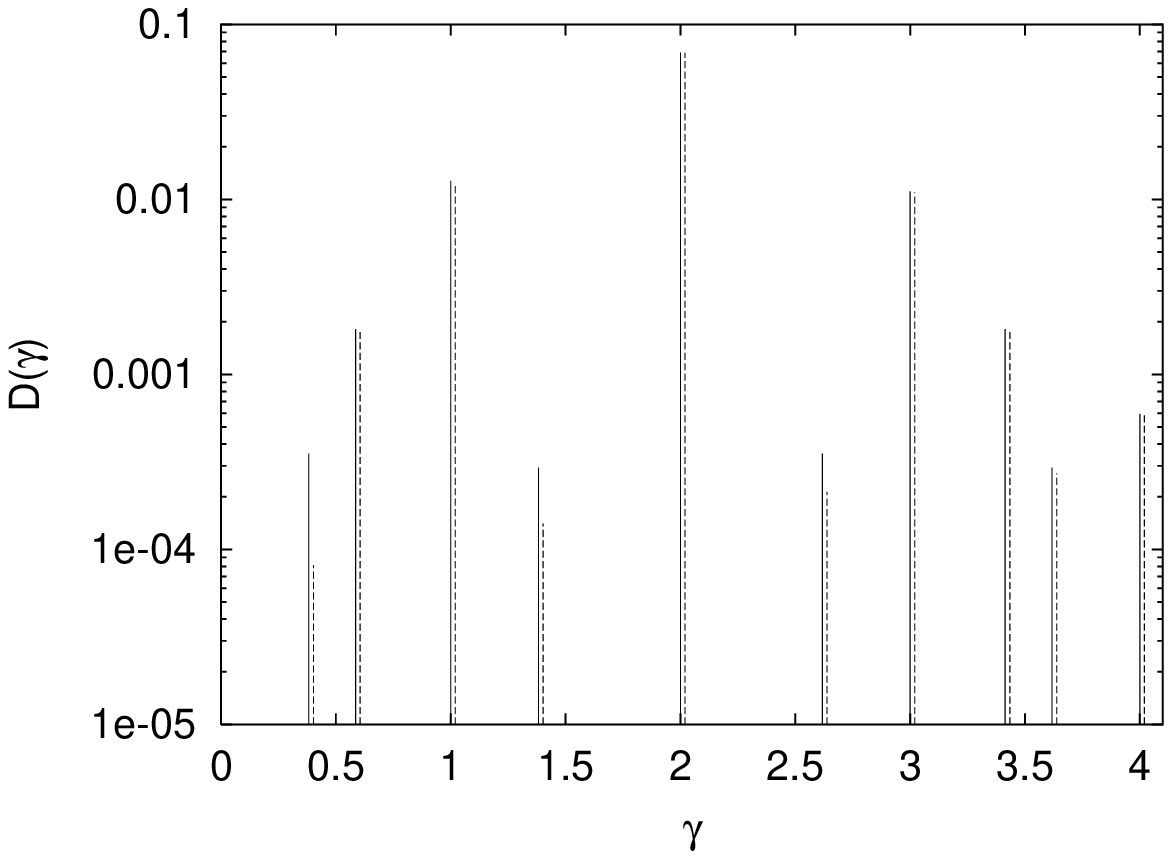}}
\end{center}
\caption{\captionCompareB}
\label{fig:compareB}
\end{figure}

\begin{figure}[htb]
\begin{center}
\myscalebox{\includegraphics{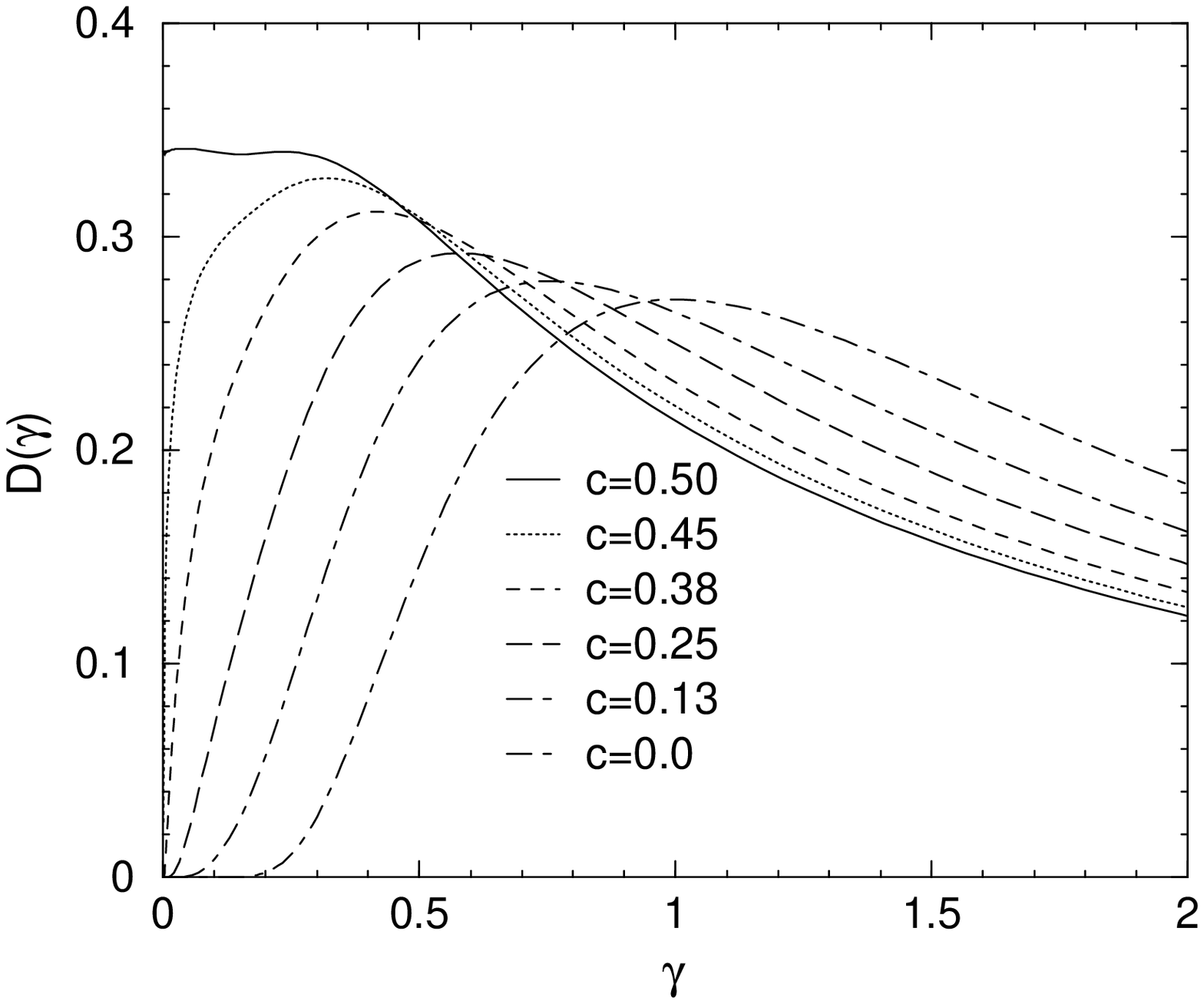}}
\end{center}
\caption{\captionFigOne}
\label{Fig1}
\end{figure}


\begin{figure}[htb]
\begin{center}
\myscalebox{\includegraphics{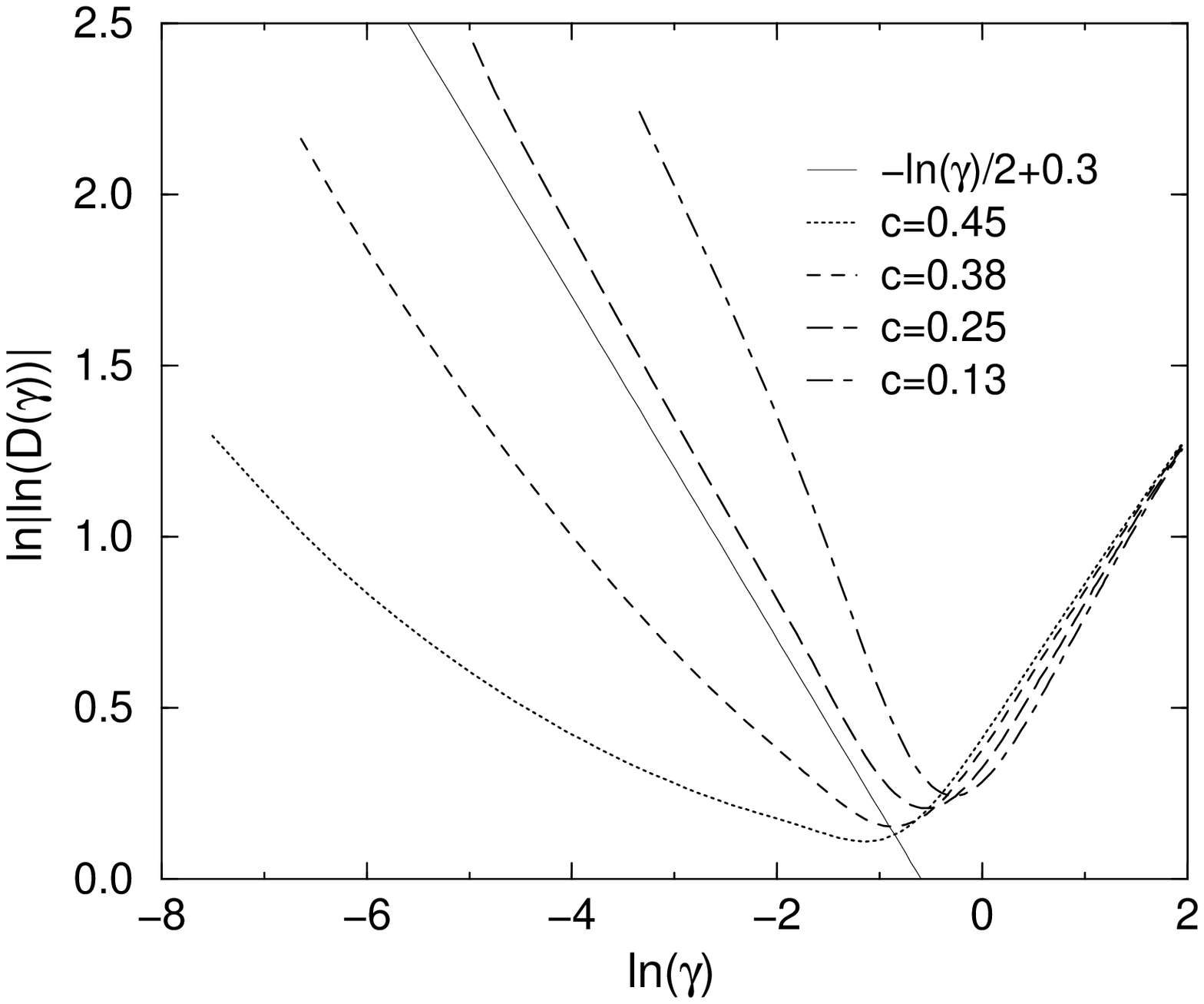}}
\end{center}
\caption{\captionFigTwo}
\label{Fig2}
\end{figure}


\begin{figure}[htb]
\begin{center}
\myscalebox{\includegraphics{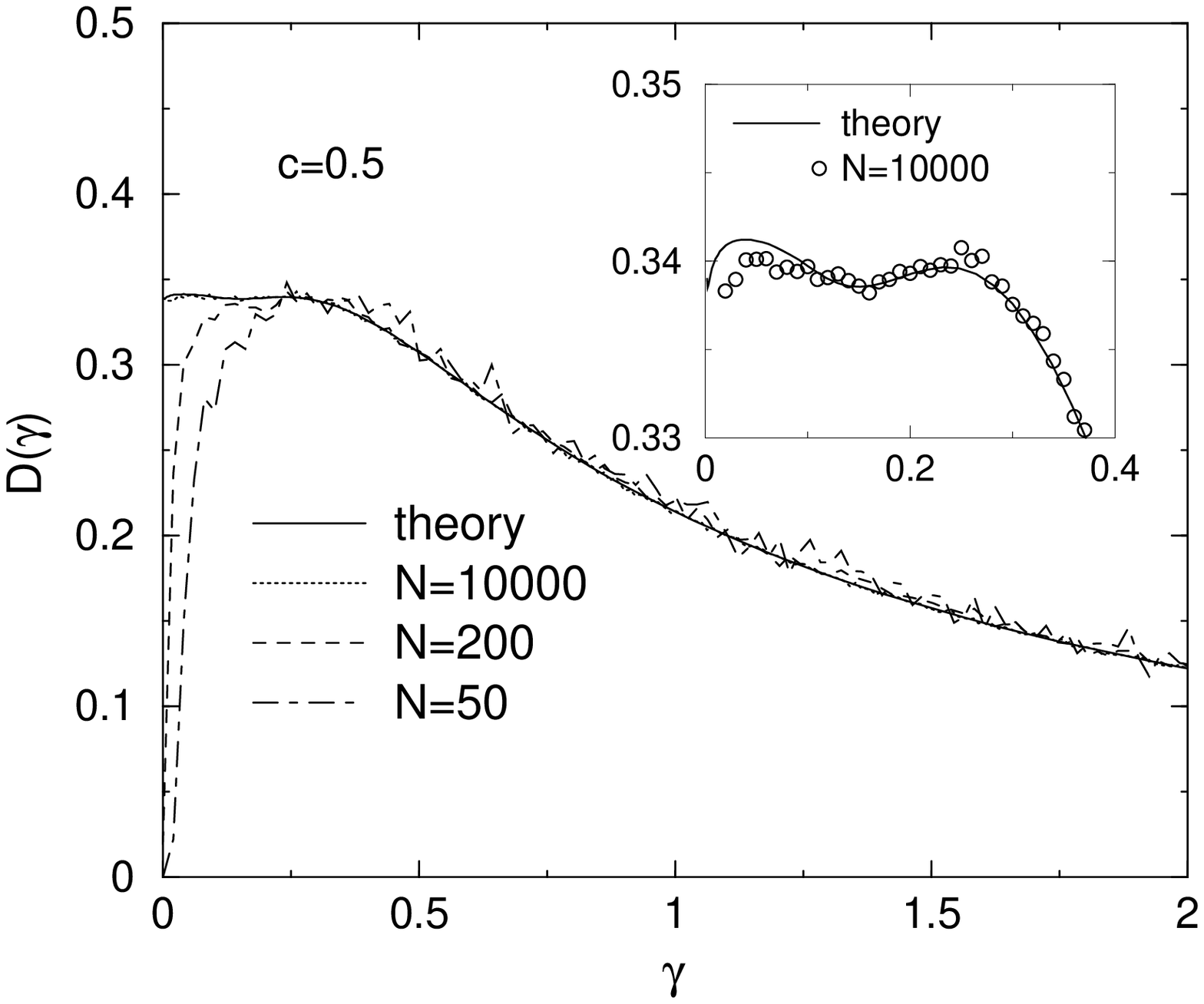}}
\end{center}
\caption{\captionDGammaCpc}
\label{figDGammaCpc}
\end{figure}

\begin{figure}[htb]
\begin{center}
\myscalebox{\includegraphics{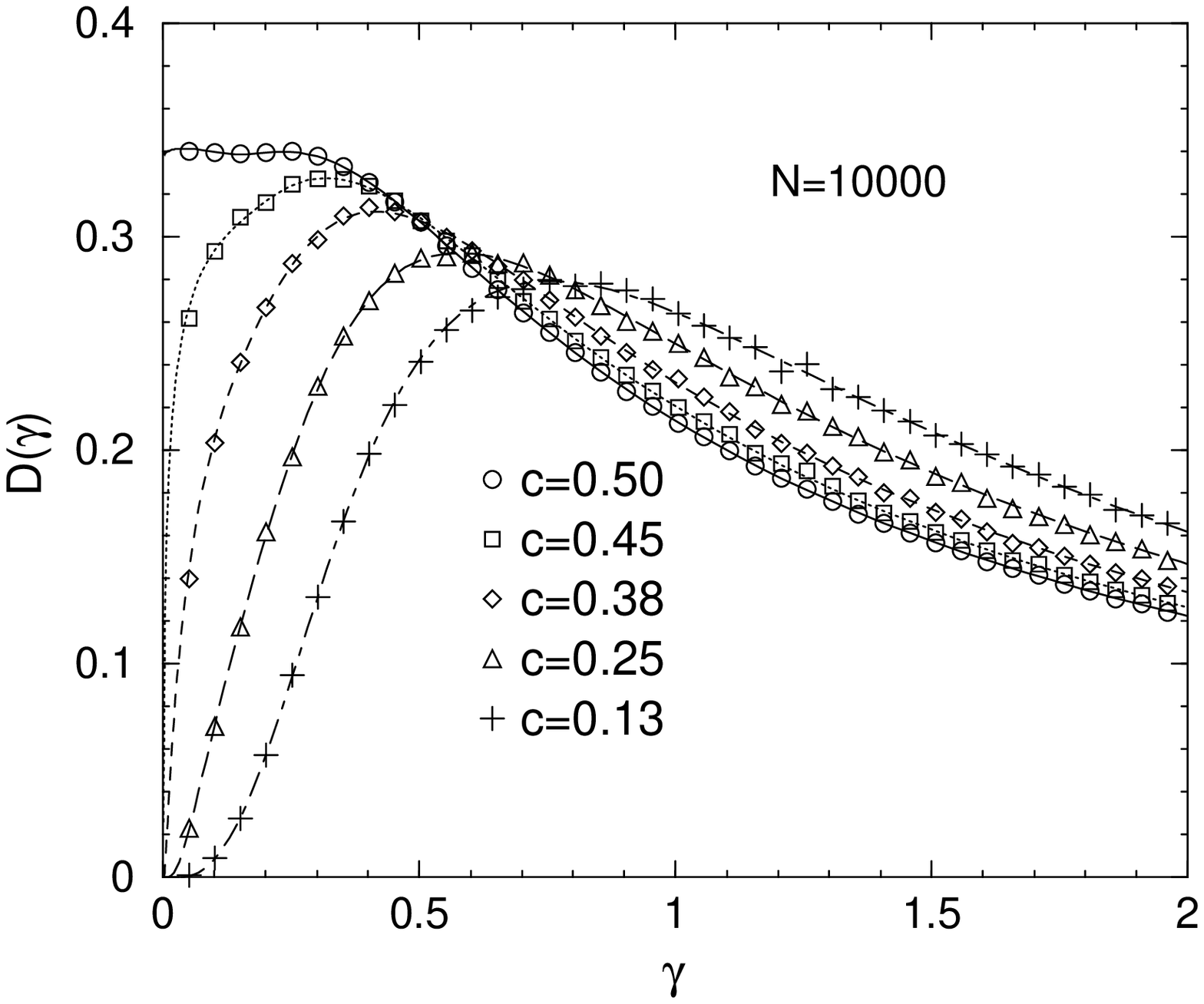}}
\end{center}
\caption{\captionDGamma}
\label{figDGamma}
\end{figure}

\begin{figure}[htb]
\begin{center}
\myscalebox{\includegraphics{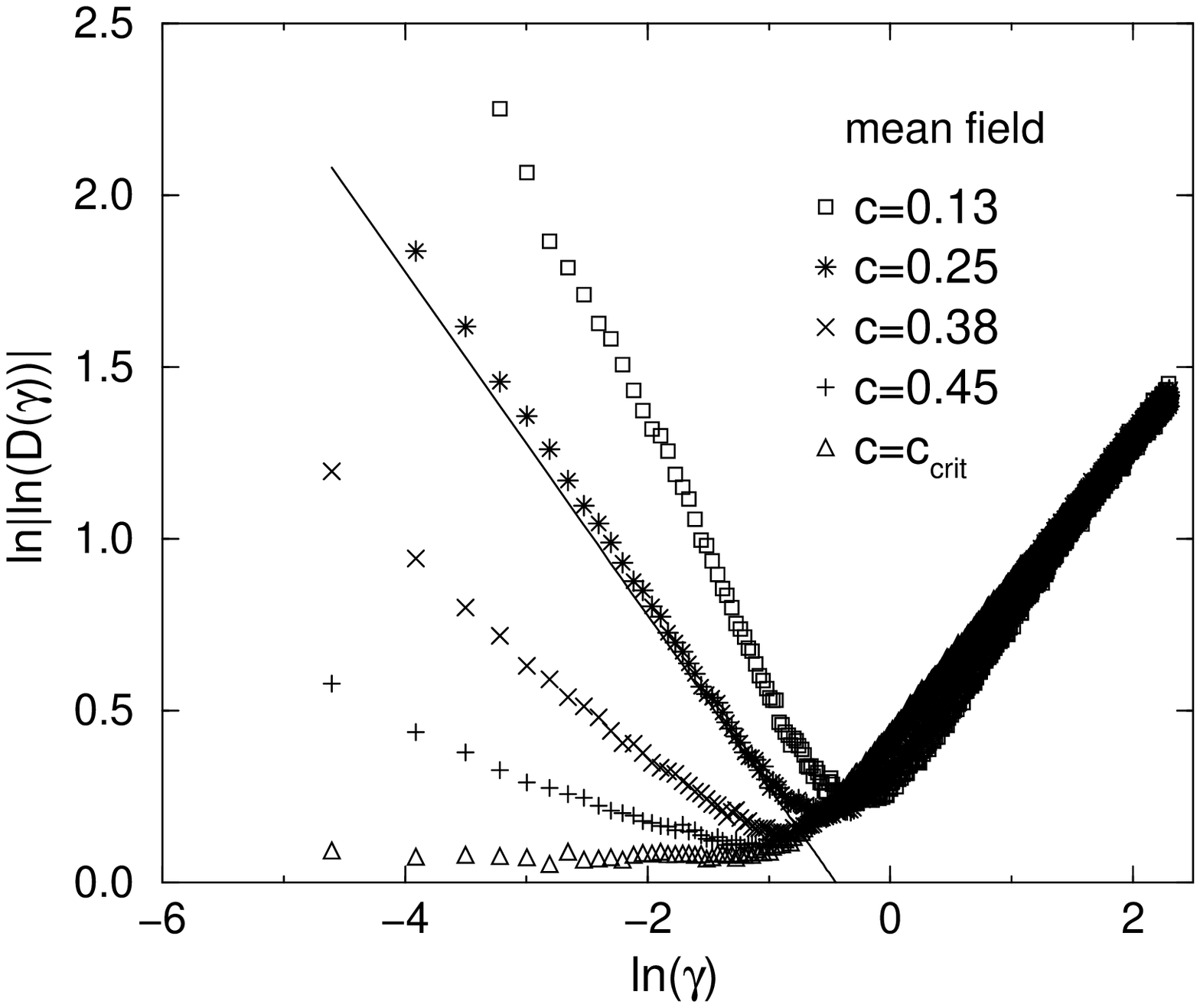}}
\end{center}
\caption{\captionErdoesLogDGamma}
\label{figErdoesLogDGamma}
\end{figure}

\begin{figure}[htb]
\begin{center}
\myscalebox{\includegraphics{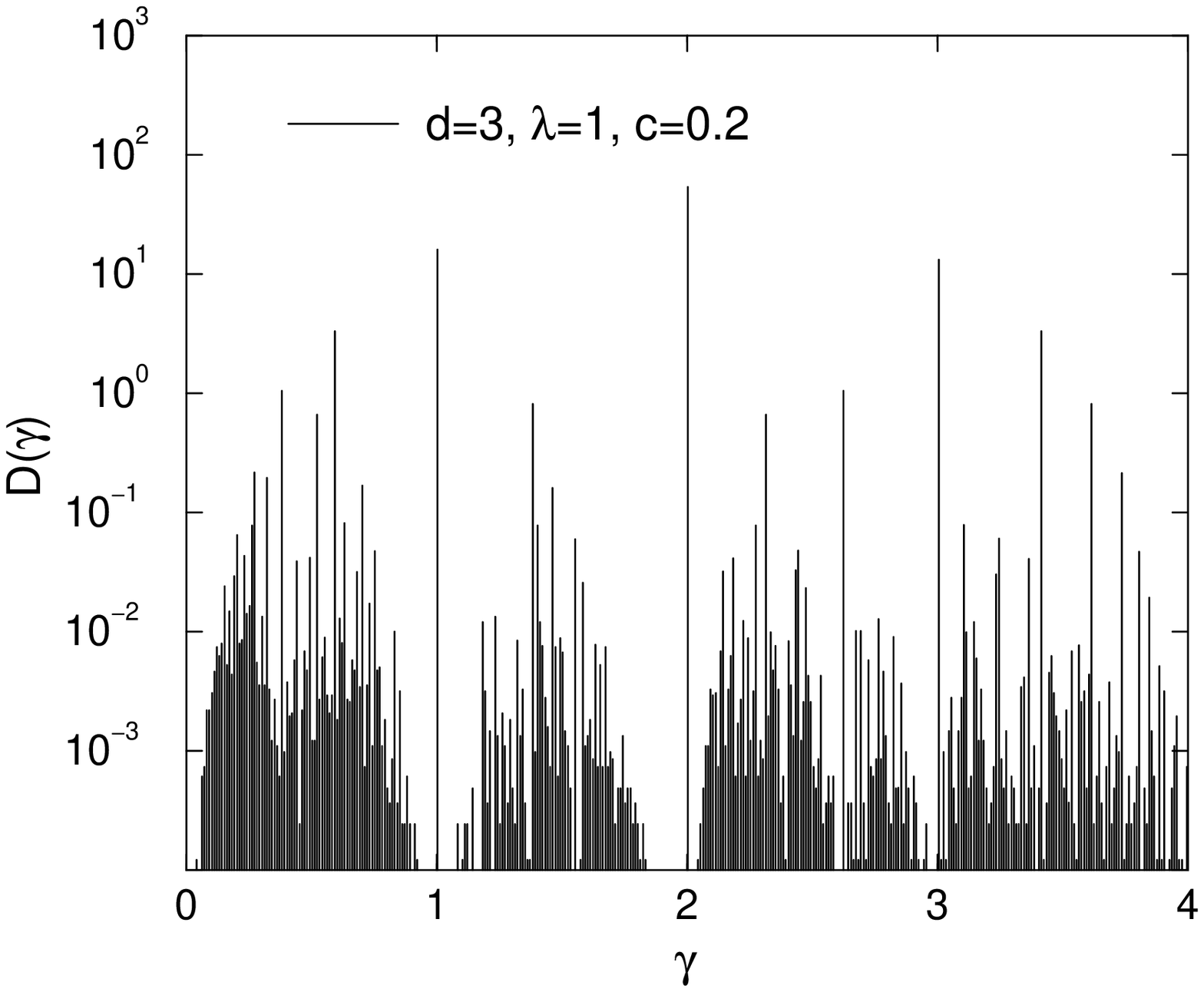}}
\end{center}
\caption{\captionGridThreeGammaDeltaDistr}
\label{figGridThreeGammaDeltaDistr}
\end{figure}

\begin{figure}[htb]
\begin{center}
\myscalebox{\includegraphics{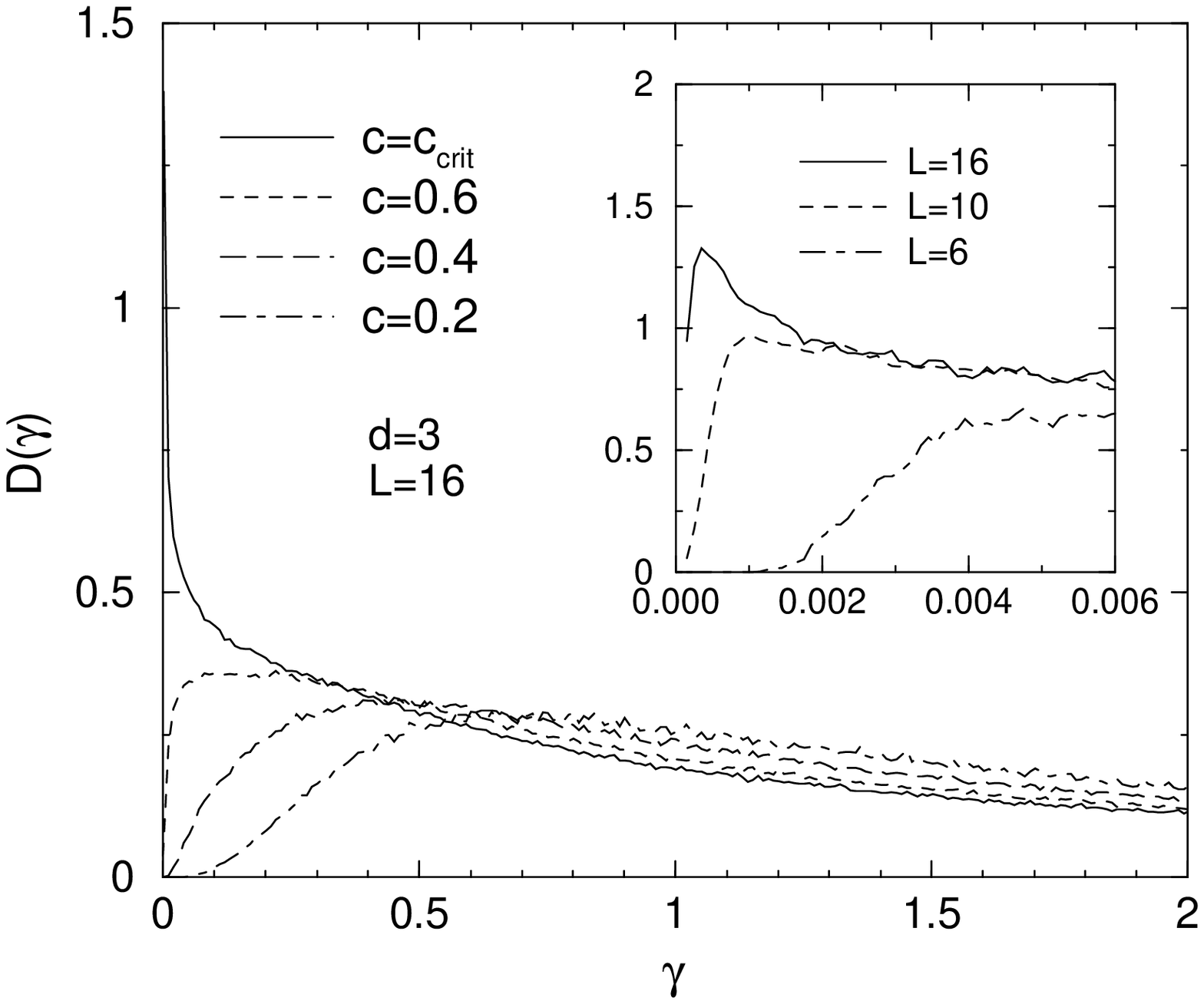}}
\end{center}
\caption{\captionGridThreeDGammaP}
\label{figGridThreeDGammaP}
\end{figure}

\begin{figure}[htb]
\begin{center}
\myscalebox{\includegraphics{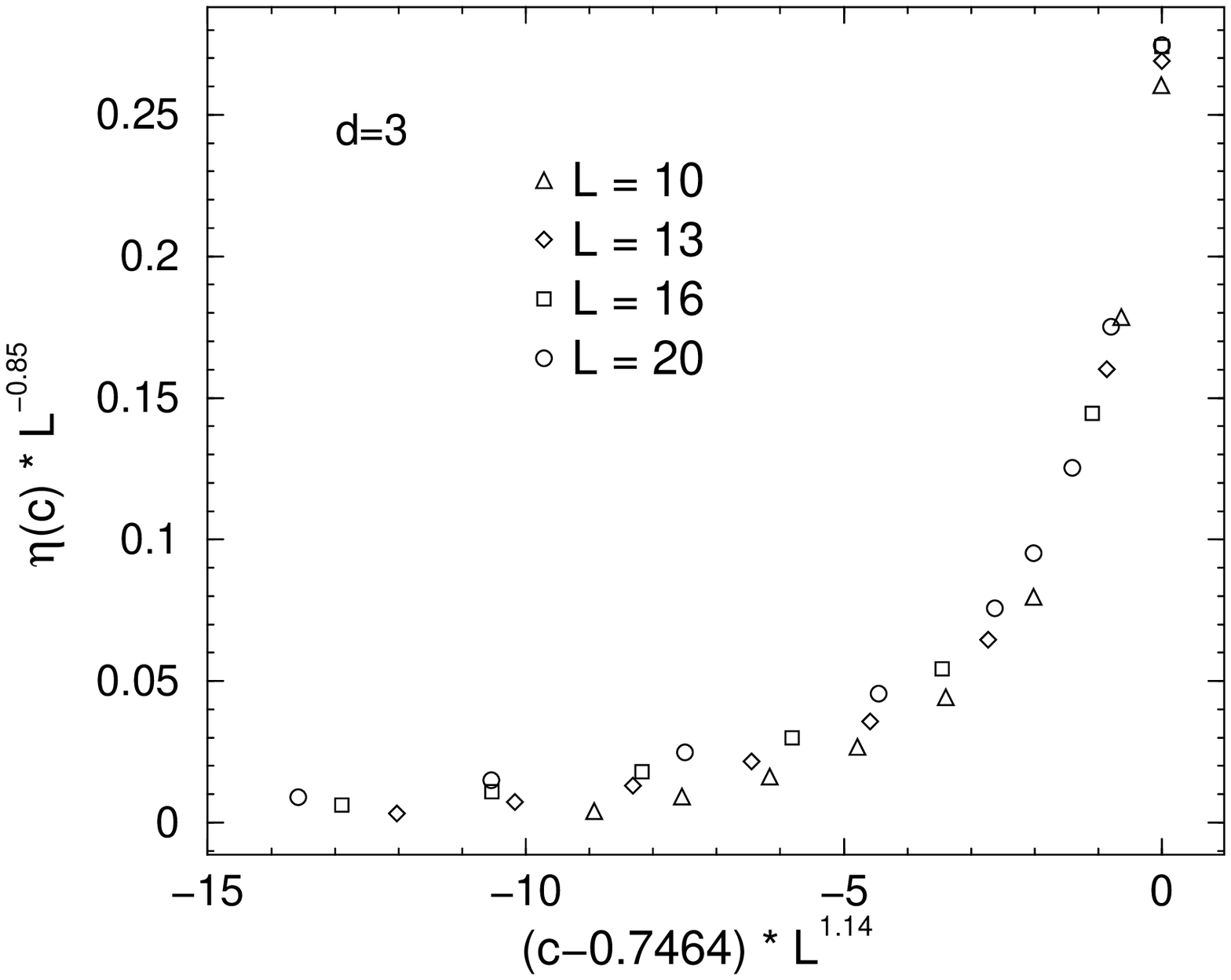}}
\end{center}
\caption{\captionViscosThreeDDelta}
\label{figViscosThreeDDelta}
\end{figure}

\begin{figure}[htb]
\begin{center}
\myscalebox{\includegraphics{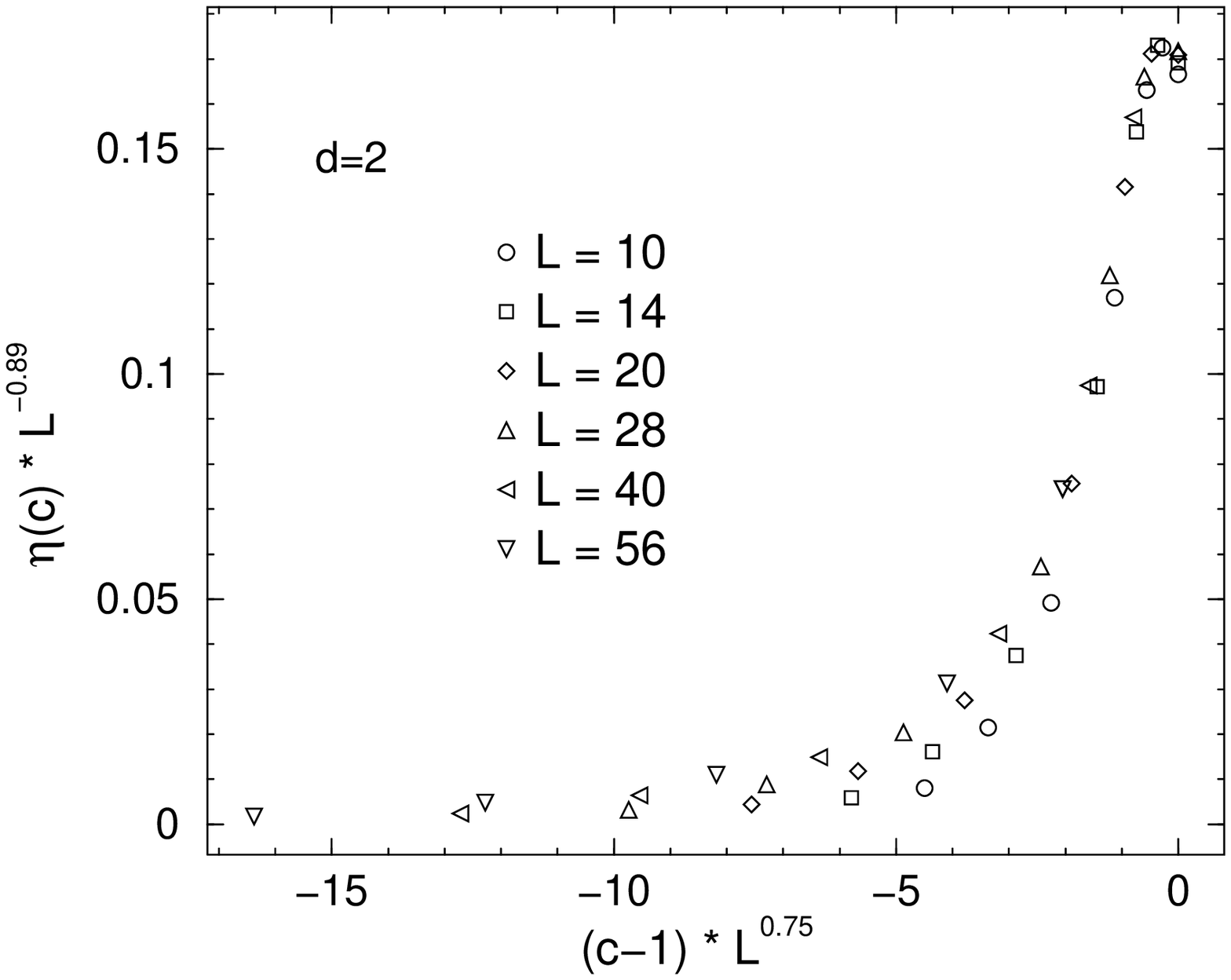}}
\end{center}
\caption{\captionViscosTwoDDelta}
\label{figViscosTwoDDelta}
\end{figure}

\begin{figure}[htb]
\begin{center}
\myscalebox{\includegraphics{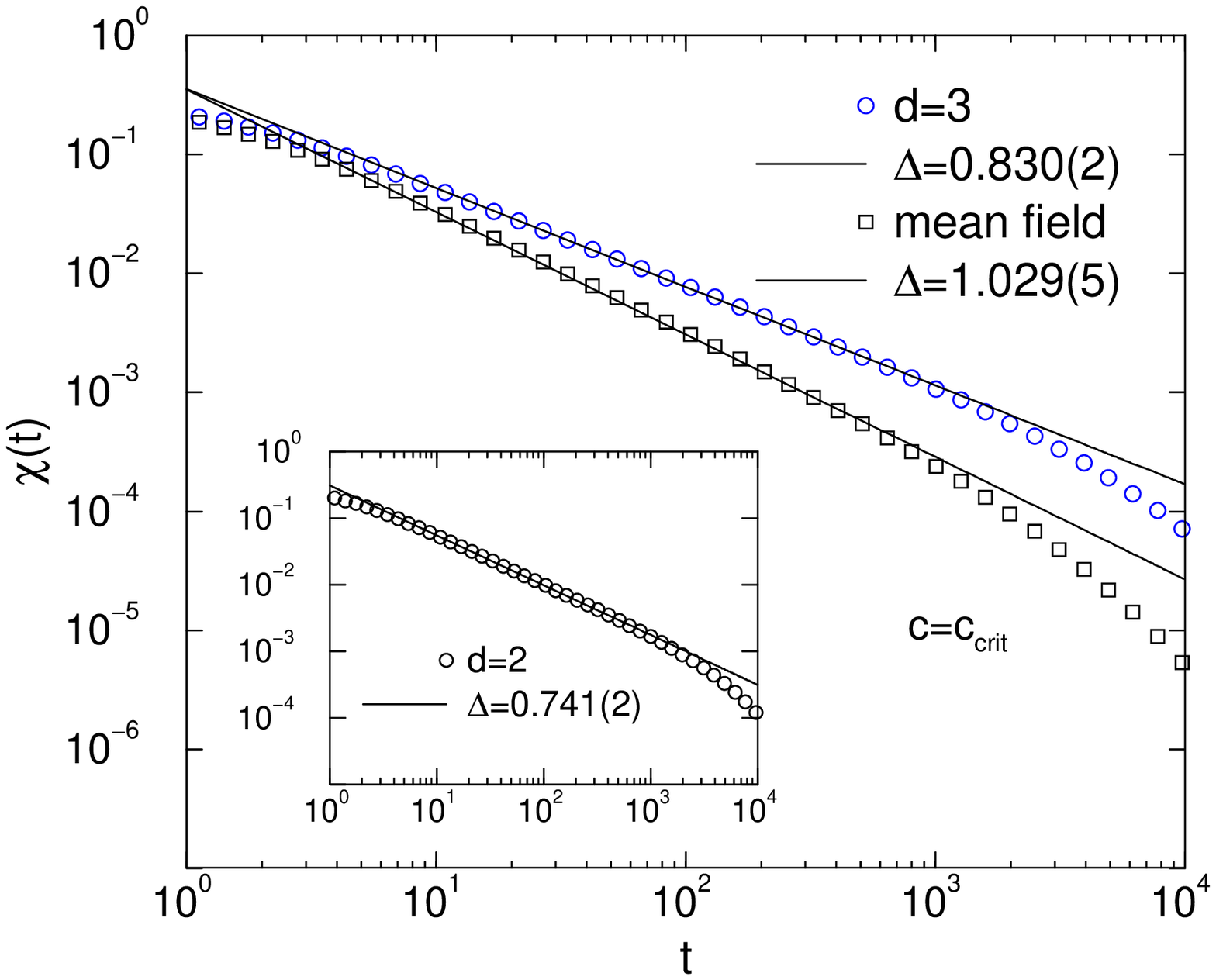}}
\end{center}
\caption{\captionGTDelta}
\label{figDTDelta}
\end{figure}

\begin{figure}[htb]
\begin{center}
\myscalebox{\includegraphics{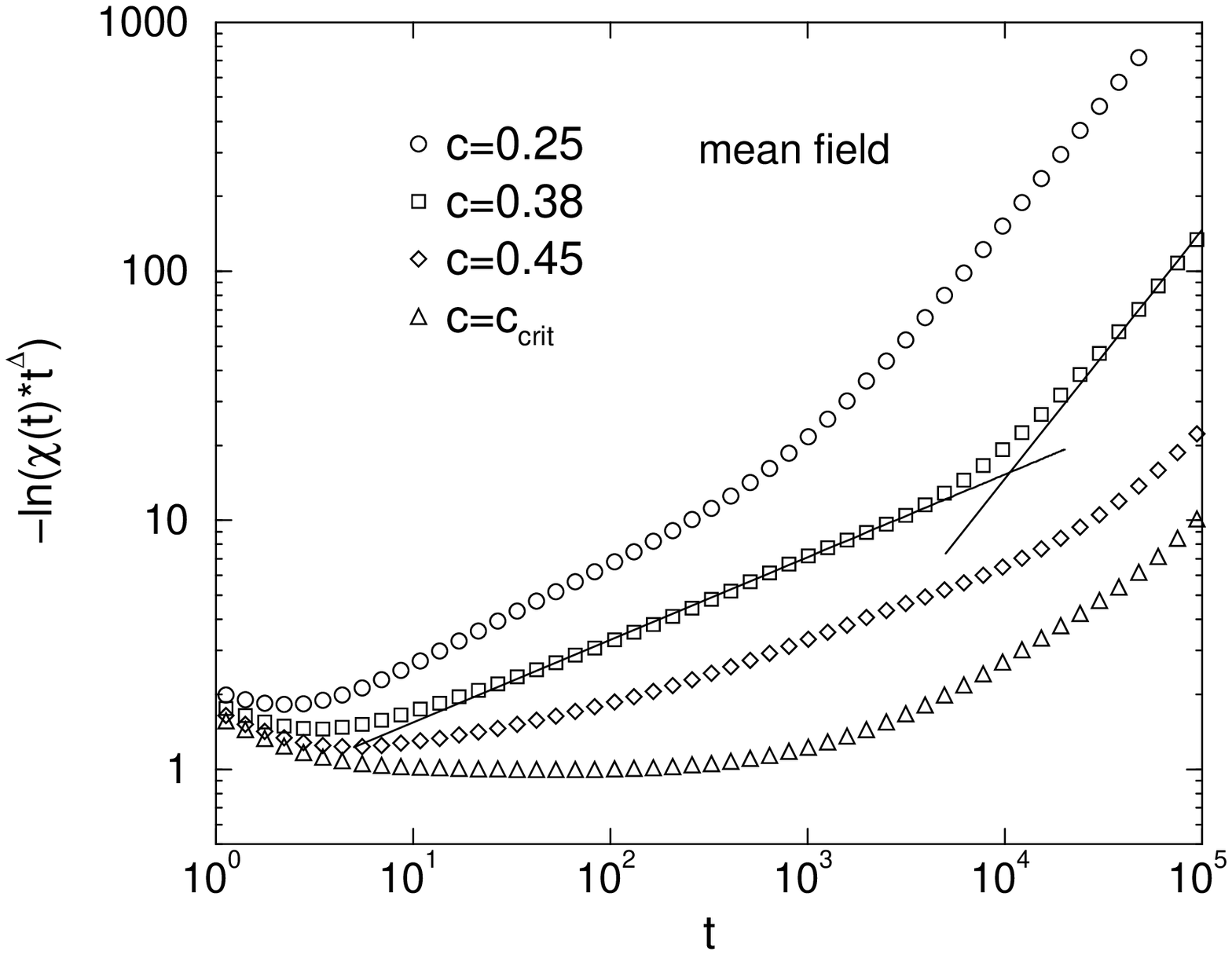}}
\end{center}
\caption{\captionGtErdoesP}
\label{figGtErdoesP}
\end{figure}

\begin{figure}[htb]
\begin{center}
\myscalebox{\includegraphics{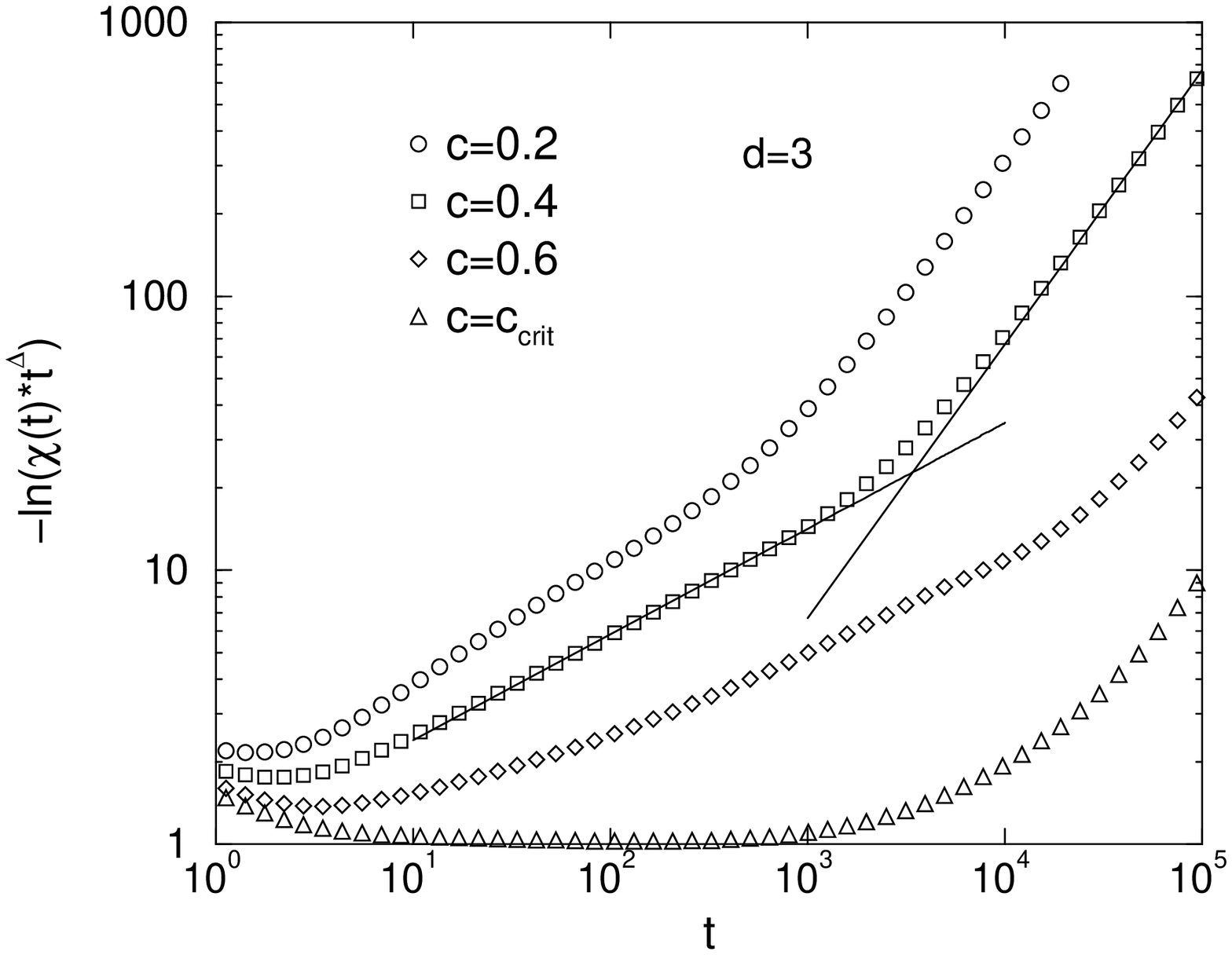}}
\end{center}
\caption{\captionGtThreeDgridP}
\label{figthreeGtDgridP}
\end{figure}

\begin{figure}[htb]
\begin{center}
\myscalebox{\includegraphics{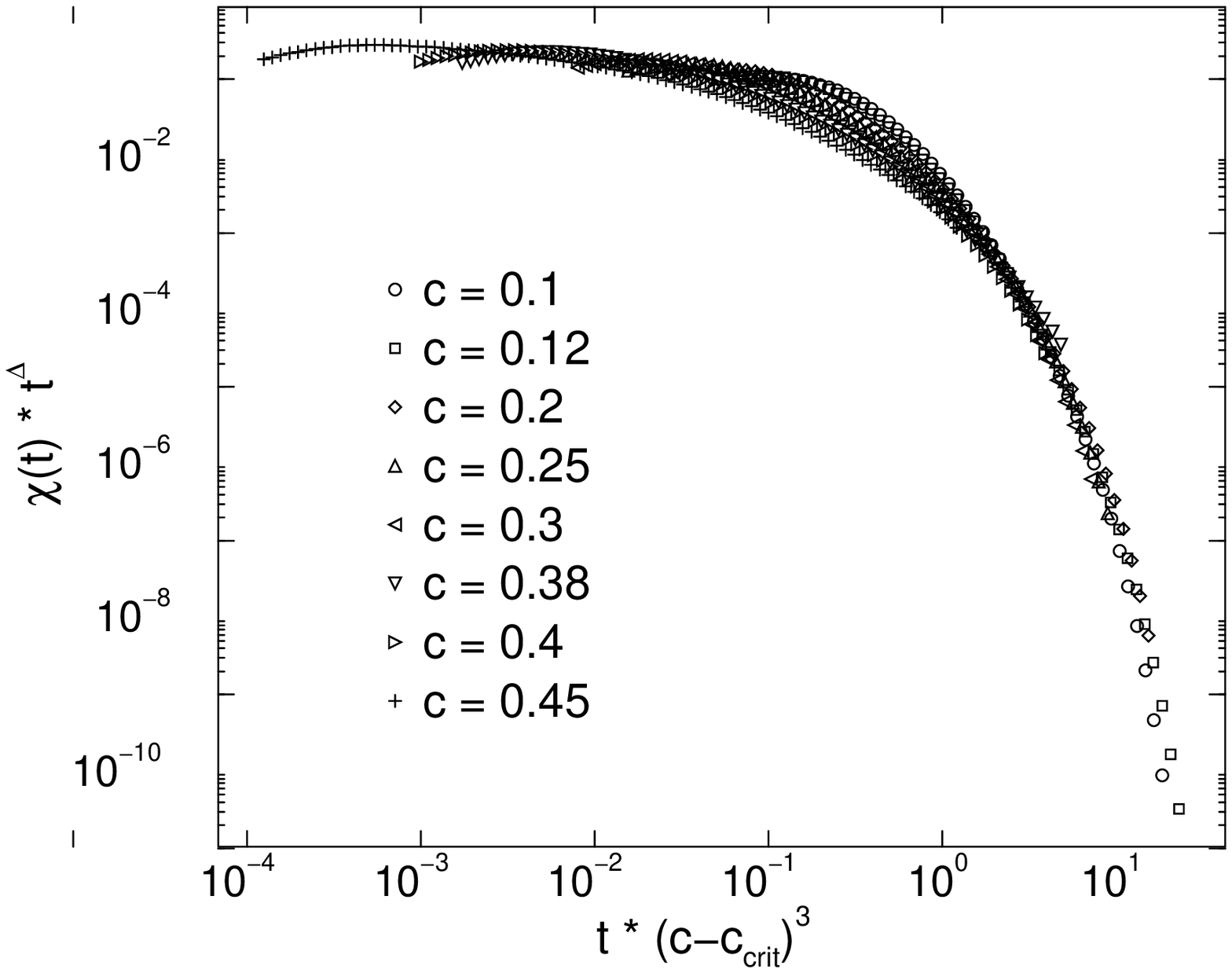}}
\end{center}
\caption{\captionGtScaleDeltaTrue}
\label{figGtScaleDeltaTrue}
\end{figure}

\end{document}